\documentclass[a4paper,fleqn,usenatbib]{mnras}

\DeclareMathAlphabet{\altmathcal}{OMS}{cmsy}{m}{n}
\usepackage{mathptmx}

\usepackage[T1]{fontenc}
\usepackage{ae,aecompl}

\usepackage{graphicx}	
\usepackage{amsmath}	
\usepackage{amssymb}	
\usepackage{xcolor}      
\usepackage{cancel}
\usepackage{multirow}
\usepackage{mathrsfs}

\usepackage[T1]{fontenc}
\usepackage{ae,aecompl}

\renewcommand{\vec}[1]{\mathbf{#1}}
\newcommand{\hvec}[1]{\hat{\boldsymbol{#1}}}
\newcommand{\pd}[2]{\frac{\partial #1}{\partial #2} }
\newcommand{\HALF}{\frac{1}{2}}

\newcommand{\av}[1]{\left< {#1} \right>}
\newcommand{\quotes}[1]{``#1''}
\newcommand{\DS}{\displaystyle}
\newcommand{\tens}[1]{\mathsf{#1}}
\newcommand{\tJ}{\tilde{\vec{J}}}
\newcommand{\cE}{\altmathcal{E}}
\newcommand{\cF}{\altmathcal{F}}
\newcommand{\cH}{\altmathcal{H}}
\newcommand{\cP}{\altmathcal{P}}

\newcommand{\cR}{\altmathcal{R}}
\newcommand{\cS}{\altmathcal{S}}

\newcommand{\cU}{\altmathcal{U}}
\newcommand{\cV}{\altmathcal{V}}
\newcommand{\cW}{\altmathcal{W}}

\newcommand{\cc}{ {\boldsymbol{c}} }
\newcommand{\cf}{ f }

\newcommand{\xf}{ {\boldsymbol{x}_f} }
\newcommand{\yf}{ {\boldsymbol{y}_f} }
\newcommand{\zf}{ {\boldsymbol{z}_f} }

\newcommand{\xe}{ {\boldsymbol{x}_e} }
\newcommand{\ye}{ {\boldsymbol{y}_e} }
\newcommand{\ze}{ {\boldsymbol{z}_e} }

\DeclareMathOperator{\sech}{sech}

\title{A Constrained Transport Method for the Solution of the Resistive
       Relativistic MHD Equations}

\author[A. Mignone]{A. Mignone$^{1,2}$\thanks{E-mail: {mignone@to.infn.it}},
 G. Mattia$^{3}$,
 G. Bodo$^{2}$,
 and L. Del Zanna$^{4,5}$
 \\
$^1$ Dipartimento di Fisica, Universit\`a di Torino, via P. Giuria 1, I-10125
Torino, Italy \\
$^2$ INAF, Osservatorio Astronomico di Torino, Strada Osservatorio 20, I-10025 Pino
Torinese, Italy \\
$^3$ Max Planck Institute for Astronomy and IMPRS -- University of Heidelberg, 
K\"onigstuhl 17, D-69117, Heidelberg Germany \\
$^4$ Dipartimento di Fisica e Astronomia, Universit\`a di Firenze e INFN -- Sez. di
Firenze, via G. Sansone 1, I-50019 Sesto F.no, Italy \\
$^5$ INAF, Osservatorio Astrofisico di Arcetri, Largo E. Fermi 5, I-50125 Firenze,
Italy }

\date{Accepted XXX. Received YYY; in original form ZZZ}

\pubyear{2019}

\begin{document}
\label{firstpage}
\pagerange{\pageref{firstpage}--\pageref{lastpage}}
\maketitle

\begin{abstract}
We describe a novel Godunov-type numerical method for solving the equations of resistive relativistic magnetohydrodynamics.
In the proposed approach, the spatial components of both magnetic and electric fields are located at zone interfaces and are evolved using the constrained transport formalism.
Direct application of Stokes' theorem to Faraday's and Ampere's laws ensures that the resulting discretization is divergence-free for the magnetic field and charge-conserving for the electric field.
Hydrodynamic variables retain, instead, the usual zone-centred representation commonly adopted in finite-volume schemes. 
Temporal discretization is based on Runge-Kutta implicit-explicit (IMEX) schemes in order to resolve the temporal scale disparity introduced by the stiff source term in Ampere's law.
The implicit step is accomplished by means of an improved and more efficient Newton-Broyden multidimensional root-finding algorithm.
The explicit step relies on a multidimensional Riemann solver to compute the line-averaged electric and magnetic fields at zone edges and it employs a one-dimensional Riemann solver at zone interfaces to update zone-centred hydrodynamic quantities. 
For the latter, we introduce a five-wave solver based on the frozen limit of the relaxation system whereby the solution to the Riemann problem can be decomposed into an outer Maxwell solver and an inner hydrodynamic solver.
A number of numerical benchmarks demonstrate that our method is superior in stability and robustness to the more popular charge-conserving divergence cleaning approach where both primary electric and magnetic fields are zone-centered.
In addition, the employment of a less diffusive Riemann solver noticeably improves the accuracy of the computations.
\end{abstract}

\begin{keywords}
methods: numerical -- relativistic processes -- MHD -- shock waves
\end{keywords}



\section{Introduction}
\label{sec:introduction}
%
%
%

The study of the dynamics of relativistic plasmas is of utmost relevance for the interpretation of the phenomenology of high energy astrophysical sources.
The ideal Magnetohydrodynamic (MHD) approximation, where dissipative processes are neglected, well describes the  large scale dynamics of a plasma and it has been extended to the relativistic regime by \citet{Lichnerowicz1967} and \citet{Anile2005}.
Over the last decade, the ideal relativistic MHD (RMHD) approach has been  used for describing the dynamics of objects like relativistic outflows and jets both from Active Galactic Nuclei (AGN) and gamma ray burst (GRB), accretion flows and  pulsar wind nebulae (PWN) by means of numerical simulations  \citep[see e.g.][]{McKinney2009,  Mimica2009, Mignone_etal2010, McKinney2012, Mukherjee_etal2013, Mizuno2015, Bromberg2016, Olmi_etal2016, Rossi2017, Bromberg2018, Bugli_etal2018} and important progresses have been reached  in establishing robust and accurate numerical schemes  for solving the ideal RMHD equations \citep[see e.g.][]{Komissarov1999, Balsara2001, dZBL2003, MB2006, Gammie2003, GR2006, dZ_etal2007, MUB2009}.
In typical astrophysical conditions, resistivity is very low  and the ideal limit is very  well suited for describing processes that occur on dynamical time scales.

However, the flow evolution may lead to the formation of localized region of large gradients 
(e.g. current sheets) where resistivity cannot be any longer neglected since its role becomes  essential in the energy and momentum balance.
Processes like magnetic reconnection can be very  important for converting magnetic energy in other forms and may play a fundamental role for interpreting the phenomenology of high energy astrophysical objects and their description requires the treatment of resistive effects.

The derivation of a consistent relativistic theory of non-ideal hydrodynamics and MHD has been achieved by several authors \citep{Lichnerowicz1967, Israel1976, Stewart1977, Carter1991, Anile2005} and in particular the equations of resistive relativistic MHD  have been derived in a relatively simple form.
More recently several authors have discussed schemes for the numerical solution of such system of equations and have presented actual implementations \citep[see, e.g.][]{Komissarov2007a, DumbserZanotti2009, Palenzuela_etal2009, TakIno2011, BdZ2013, Dionys_etal2013, Mizuno2013, Miranda_etal2018}.

From the numerical point of view, the solution of the resistive RMHD equations is more challenging than approaching their ideal counterpart as it draws on two main issues. 
First, the resistive RMHD equations are hyperbolic with a stiff relaxation term which accounts for the large difference between the dynamical and the diffusive time scales posing very strict constraints on the time step used for time-explicit calculations.
One possibility for overcoming this problem is offered by implicit-explicit Runge-Kutta schemes (IMEX) \citep{Pareschi2005}.
Such an approach represents a very effective solution to the problem, by combining the simplicity of an explicit treatment of the flux avoiding the time-step restrictions due to the stiffness.
IMEX schemes for the resistive RMHD have been implemented and tested by several of the previously mentioned authors \citep[e.g.][]{Palenzuela_etal2009, BdZ2013, Dionys_etal2013, Miranda_etal2018}, showing that they are very well suited for such system of equations. The second issue is related to the solenoidal condition and charge conservation.
From an analytical point of view, both are direct consequences of Maxwell's equations; at the discrete level, however, this does no longer hold because of the numerical errors introduced by the underlying algorithm.
More specifically, the numerical counterparts of the divergence and the curl operator do not ensure that $\nabla\cdot(\nabla\times\vec{A}) = 0$, where $\vec{A}$ is any vector field.
As a consequence, for the magnetic field, the constraint $\nabla\cdot\vec{B} = 0$  may not be maintained during the evolution while, for the electric field, it is the condition $\nabla\cdot\vec{E} = q$ that is not respected (when $\vec{E}$ and $q$ are evolved through Ampere's law and the charge conservation equation, respectively).
This problem turns out to be particularly severe in shock-capturing schemes, where the stencils used during the reconstruction routines and the related accuracy may vary along the different directions, with the consequence that numerical partial derivatives do not commute \citep{LdZ2004}.

Most of the previously mentioned investigators have devised numerical schemes based on a divergence cleaning approach whereby one solves a modified system of conservation laws where Faraday's and Ampere's law are coupled to generalized Lagrange multipliers \citep[GLM,][]{Munz_etal2000, Dedner_etal2002, MigTze2010}.
Divergence errors are convected out of the computational domain at the maximum characteristic speed and damped at the same time.
The GLM approach offers ease of implementation since fluid variables and electromagnetic field retain a zone-centered representation.
Conversely, in the constrained transport (CT) method originally introduced by \cite{EvansHawley1988} \citep[see also][in the context of Godunov-type MHD schemes]{Balsara_Spicer1999, LdZ2004}, the magnetic field has a staggered representation whereby the different components live on the face they are normal to.
The numerical representation of the divergence and curl operators ensures that the condition $\nabla\cdot(\nabla\times\vec{E}) = 0$ is verified at discrete level, thus maintaining  $\nabla\cdot\vec{B} = 0$ to machine accuracy during the evolution.
This approach has been used by \cite{BdZ2013} \citep[see also][for a genuinely third-order scheme]{dZBB2014}, in the context of resistive RMHD, to evolve the magnetic field while still keeping a zone-centered discretization for the electric field.

In the present work, we extend the constrained transport formalism also to the electric field and follow an approach similar to that outlined in \cite{Balsara_etal2016}, in the context of the two-fluid equations.
There, an alternative staggered collocation for the electric field has been introduced that is both compatible with a Godunov scheme and gives an update of the Amp\'ere's law consistent with Gauss's law.
In the proposed method, the primary electric and magnetic field variables share the same staggered representation and are thus represented by their surface averages.
Hydrodynamic quantities retain instead the usual zone-centered collocation and are interpreted as volume averages.
The resistive RMHD equation solver has been implemented as part of the PLUTO code for astrophysical gas dynamics \citep{Mignone_PLUTO2007} and includes both the standard GLM schemes as well as the newly proposed CT scheme.

The paper is structured as follows.
In Section \ref{sec:equations} we review the fundamental equations of resistive RMHD, starting from their covariant form.
In Section \ref{sec:method_ct} the new constrained transport formulation is presented while our five-wave resistive RMHD Riemann solver is derived in Section \ref{sec:method_riemann}.
Numerical benchmarks are presented in Section \ref{sec:tests} and conclusions are finally drawn in Section \ref{sec:summary}.
 
\section{Equations}
\label{sec:equations}
%
%
%

In the present section we describe the equations for resistive relativistic MHD, first in general covariant form and later specialized to a Minkowski flat spacetime, separating time and space components and derivatives as needed for practical implementation in a numerical scheme.
In the following we will adopt physical units where $c=4\pi = 1$, a signature $(-1,+1,+1,+1)$, $g_{\mu\nu}$ will be the metric tensor, and $\nabla_\mu$ the covariant derivative associated to the metric. We will use Greek letters for covariant four-dimensional components and Latin letters for spatial three-dimensional ones.

\subsection{Covariant formalism}
%

The equations of relativistic MHD are composed by a first couple of conservation laws, one for baryon number (or equivalently mass, assuming a single fluid with particles of given rest mass) and one for total momentum-energy conservation:
\begin{equation}\label{eq:hydro}
  \nabla_\mu (\rho u^\mu)  =  0, \qquad \nabla_\mu T_\mathrm{tot}^{\mu\nu}  =  0,
\end{equation}
where $\rho$ is the rest mass density, $u^\mu$ the fluid four-velocity, and $T^{\mu\nu}_\mathrm{tot}$ the total (matter and fields) stress-energy tensor.
The second couple is that of Maxwell's equations
\begin{equation}\label{eq:maxwell}
  \nabla_\mu F^{\mu\nu} = - J^\nu, \quad \nabla_\mu F^{\star\mu\nu} = 0,
\end{equation}
where $F^{\mu\nu}$ is the Faraday electromagnetic (EM) tensor, $F^{\star\mu\nu}$ its dual, and $J^\mu$ the four-current density, which, due to the anti-symmetric property, satisfies the condition 
\begin{equation}\label{eq:current}
\nabla_\mu J^\mu = 0
\end{equation}
of electric charge conservation. 
Let us now split the total stress-energy tensor into the gas and electromagnetic field (EM) components, for which we have
\begin{equation}\label{eq:mf}
     \nabla_\mu T^{\mu\nu}_{\rm g}
  = -\nabla_\mu T^{\mu\nu}_{\rm EM} = - J_\mu\,F^{\mu\nu},
\end{equation}
where the last term is the Lorentz force acting on the charged fluid.

We then decompose our quantities according to $u^\mu$.
In the case of an ideal fluid, the matter contribution to the energy-momentum tensor is simply provided by ideal hydrodynamics as
\begin{equation}
  T_{\rm g}^{\mu\nu} = (\varepsilon + p) u^\mu u^\nu + p \, g^{\mu\nu},
\end{equation}
where $\varepsilon = T_{\rm g}^{\mu\nu} u_\mu u_\nu$ is the gas energy density, and $p$ is the kinetic pressure.
The EM tensor and its dual can be expressed as
\begin{align}
  F^{\mu\nu} &  =   u^\mu e^\nu - u^\nu e^\mu
                         + \epsilon^{\mu\nu\lambda\kappa} b_\lambda u_\kappa, \nonumber \\
  F^{\star\mu\nu} & =   u^\mu b^\nu - u^\nu b^\mu
                             - \epsilon^{\mu\nu\lambda\kappa} e_\lambda u_\kappa,
\label{eq:f1}
\end{align}
where $\epsilon^{\mu\nu\lambda\kappa}$ is the Levi-Civita pseudo-tensor, and where the vectors
\begin{equation}\label{eq:ebdef}
  e^\mu = F^{\mu\nu}u_\nu, \qquad b^\mu = F^{\star\mu\nu}u_\nu =
\tfrac{1}{2} \epsilon^{\mu\nu\lambda\kappa} F_{\lambda\kappa} u_\nu ,
\end{equation}
are the electric and magnetic fields measured in the fluid rest frame.
Given these definitions, the field component of the stress-energy tensor can be written as
\begin{equation}\label{eq:tf}
  T_{\rm EM}^{\mu\nu} = (e^2 + b^2)u^\mu u^\nu
                              + \tfrac{1}{2}(e^2+b^2)g^{\mu\nu}
                              - e^\mu e^\nu - b^\mu b^\nu.
\end{equation}
The four-current can be also split according to
\begin{equation}
  J^\mu = q_0 u^\mu + j^\mu,
\end{equation}
where $q_0 = - J^\mu u_\mu$ is the proper electric charge density and $j^\mu$ the \emph{conduction} current density.
We show in Appendix \ref{app:q0} that the rest-frame charge density can be  expressed in terms of the comoving fields and the kinematic vorticity (see Eq. \ref{eq:q0}).
We also point out that even in the ideal limit ($\eta\to0$) the rest-frame charge does not vanish but it can be expressed as $q_0 = -b^\mu\omega_\mu$, where $\omega_\mu$ is the kinematic vorticity.

Projecting the momentum-energy conservation law Eq.~(\ref{eq:mf}) across the flow yields the equation of motion
\begin{equation}
  (\varepsilon + p) u^\nu \nabla_\nu  u_\mu + \nabla_\mu p + u_\mu u^\nu \nabla_\nu p = 
  q_0 e_\mu + \epsilon_{\mu\nu\lambda\kappa} j^\nu b^\lambda u^\kappa,
\end{equation}
in which the right hand side is the Lorentz force, whereas the energy equation is obtained by projecting along the flow, so that
\begin{equation}
  u^\mu\nabla_\mu\varepsilon + (\varepsilon + p) \nabla_\mu u^\mu = j_\mu e^\mu,
\end{equation}
where Joule heating acts as an energy source. We clearly need some sort of Ohm's law to close the system and to specify the heating term.

In the literature three possibilities are most commonly adopted:
\begin{enumerate}

\item 
\emph{ideal plasma} -- the mobility of charge carriers is so high that the comoving electric field must vanish in order to prevent huge currents.
Therefore Ohm's law is
\begin{equation}
  e^\mu = 0,
\end{equation}
and only the source-less Maxwell equation is needed to be solved (for $b^\mu$);

\item
\emph{resistive plasma} -- we assume an isotropic tensor of electric conductivity, so that
Ohm's law is simply
\begin{equation}\label{eq:ohm0}
  e^\mu = \eta j^\mu,
\end{equation}
where $\eta$ is the (scalar) resistivity coefficient, and Joule heating retains the usual form $\eta j^2$;

\item
\emph{dynamo-chiral resistive plasma} -- in addition to resistive dissipative effects,
\emph{mean-field dynamo} or \emph{chiral magnetic effects} (CME) may be at work
in the plasma, leading to an additional current component along the magnetic field
and to magnetic field amplification.
In \cite{dZB2018} the following covariant form of Ohm's law was proposed
\begin{equation}
  j^\mu = \sigma_E e^\mu + \sigma_B b^\mu,
\end{equation}
where subscripts have been added to distinguish the Ohmic and dynamo/chiral effects, coupling to $e^\mu$ and $b^\mu$, respectively. For $\sigma_B = 0$ we retrieve the previous case, where $\sigma_E= 1/\eta$ is the conduction coefficient.
\end{enumerate}
From now on only the second option will be discussed, leaving the implementation of the dynamo/chiral case to a future work. 

\subsection{The equations for a flat spacetime}
\label{sec:eq_flatmetric}
%

The equations are now rewritten for a flat Minkowski spacetime, that is for special resistive RMHD.
We are going to separate time and space components and from now on we will use boldface notation and Latin indices for spatial vectors.
The fluid four-velocity is now written as
\begin{equation}
  u^\mu = ( \gamma, \vec{u}),
\end{equation}
where $\gamma = \sqrt{1 + u^2}$ is the Lorentz factor and $\vec{v} = \vec{u}/\gamma$ is the usual three-velocity.
The four-current is now split as
\begin{equation}
   J^\mu = (q, \vec{J}), 
\end{equation}
where $q$ is the charge density (now measured in the laboratory frame), and $\vec{J}$ the usual three-current.
The EM fields are derived from the Faraday tensor and its dual as
\begin{align}
  e^\mu & = (\vec{u}\cdot\vec{E},\, \gamma \vec{E} + \vec{u}\times\vec{B}), \\
  b^\mu & = (\vec{u}\cdot\vec{B},\, \gamma \vec{B} - \vec{u}\times\vec{E}),
\end{align}
in which $\vec{E}$ and $\vec{B}$ are the electric and magnetic field measured in the laboratory frame.

The set of resistive relativistic equations arising from the time and space split of the covariant Eqs.~(\ref{eq:hydro}) and (\ref{eq:maxwell}) are, in vectorial form,
\begin{equation}\label{eq:ResRMHD}
  \begin{array}{ll}
  &\DS \pd{D}{t} + \vec{\nabla}\cdot ( D \vec{v}) = 0,
  \\  \noalign{\medskip}
  &\DS \pd{\vec{m}}{t} + \vec{\nabla}\cdot (w \vec{u}\vec{u} + p \tens{I} + \tens{T}) = 0,
  \\  \noalign{\medskip}
  &\DS \pd{\cE}{t} + \vec{\nabla}\cdot\vec{m} = 0,
  \\  \noalign{\medskip}
  &\DS \pd{\vec{B}}{t}  + \vec{\nabla}\times\vec{E} = 0,
  \\  \noalign{\medskip}
  &\DS \pd{\vec{E}}{t}  - \vec{\nabla}\times\vec{B} = - \vec{J}, 
  \end{array}
\end{equation}
where $\tens{I}$ is the identity matrix and the fluid conserved variables are the density $D=\rho\gamma$ as measured in the laboratory frame, the total momentum density $\vec{m}= w \gamma \vec{u} + \vec{E}\times\vec{B}$, and the total energy density
\begin{equation}\label{eq:totE}
  \mathcal{E} = w \gamma^2 -  p + \cP_{\rm EM}\,.             
\end{equation}
In the expressions above, $w=\varepsilon + p$ is the specific enthalpy and $\cP_{\rm EM} = (E^2 + B^2)/2$ denotes the EM energy density.
Finally, 
\begin{equation}\label{eq:MaxwellStress}
   \tens{T} = -  \vec{E}\vec{E} - \vec{B}\vec{B}  + \tfrac{1}{2}(E^2 + B^2) \tens{I}
\end{equation}
is the Maxwell's stress tensor.
The remaining Maxwell's equations give the constraints
\begin{equation}
  \vec{\nabla}\cdot\vec{B}=0, \qquad 
  \vec{\nabla}\cdot\vec{E}=q,
\end{equation}
and the charge and current densities are also bound to satisfy the conservation equation
\begin{equation}
\DS \pd{q}{t} + \vec{\nabla}\cdot \vec{J} = 0 ,
\end{equation}
which directly follows from Eq. (\ref{eq:current}).
 
These quantities both enter in Ohm's law, Eq. (\ref{eq:ohm0}), here rewritten in terms of spatial vectors alone.
From its time component one can derive the $q_0\gamma$ term, so that the Ohm's law for the spatial current becomes
\begin{equation}\label{eq:ohm}
  \vec{J} = \frac{1}{\eta}\Big[\gamma\vec{E} + \vec{u}\times\vec{B}
                               - (\vec{E}\cdot\vec{u})\vec{v}\Big] + q\vec{v},
\end{equation}
where $q=\vec{\nabla}\cdot\vec{E}$ from Gauss' law, so that the current is determined once the fluid velocity and the electromagnetic fields are known, for a given value of the resistivity $\eta$. 
In the ideal MHD limit, $\eta\to 0$, the condition $\vec{E} + \vec{v}\times\vec{B}=0$ is retrieved (here $\vec{E}$ can be considered a secondary variable with respect to $\vec{v}$ and $\vec{B}$, to be derived by the above condition), whereas in the resistive limit, $\eta\to\infty$, we have $\vec{J}=q\vec{v}$, and the charge density $q$ satisfies a continuity equation.

For numerical purposes, we use more compact notations and rewrite the system in quasi-conservative form as
\begin{equation}\label{eq:cons_law}
 \pd{\cU}{t} = -\nabla\cdot\tens{F}(\cU) + \cS_e + \frac{1}{\eta}\cS(\cU)
             = \cR(\cU) + \frac{1}{\eta}\cS(\cU) ,
\end{equation}
where $\cU = (D,m_i,{\cal E}, B_i,E_i)^\intercal$ is the full array of conserved quantities, $\tens{F}$ is the flux tensor 
\begin{equation}\label{eq:fluxTensor}
 \tens{F} = \left( \begin{array}{c}
  \rho u_j \\
  w u_i u_j + p \delta_{ij} + \tens{T}_{ij} \\
   m_j \\
   \varepsilon^{ijk} E_k \\
  -\varepsilon^{ijk} B_k
 \end{array}\right)^\intercal ,
\end{equation}
where $\varepsilon^{ijk}$ is the three-dimensional Levi-Civita symbol.
The source term is non-zero only in the Ampere's law and it contains the current density $\vec{J}$ (Eq. \ref{eq:ohm}) which we split, for computational purposes, into a stiff ($\tJ$) and non-stiff ($q\vec{v}$) contribution.
The source terms $\cS$ and $\cS_e$ in Eq. (\ref{eq:cons_law}) take thus the form
\begin{equation}
 \cS_e =  \left(\begin{array}{c}
          0_{\times8} \\
          q\vec{v}
        \end{array}\right)
 \,,\quad
  \frac{1}{\eta}\cS = \left( \begin{array}{c}
    0_{\times8} \\
    \tJ
   \end{array}\right) .
\end{equation}

In \cite{MMB2018} it has been shown that the system of hyperbolic PDE given by (\ref{eq:cons_law}) admits 10 propagating modes which are easily recognized in the limits of small or large conductivities.
In the $\eta\to\infty$ limit, matter and electromagnetic fields decouple and solution modes approach pairs of light and acoustic waves as well as a number of purely damped (non-propagating) modes.
In the $\eta\to0$ (ideal) limit, modes of propagation coincide with a pair of fast magnetosonic, a pair of slow and Alfv\'en modes, as expected.
The contact mode is always present and it is unaffected by the conductivity.

\section{Description of the CT-IMEX Scheme}
\label{sec:method_ct}
%
%
%

\subsection{Notations and General Formalism}
\label{sec:notations}
%

We adopt a Cartesian coordinate system with unit vectors $\hvec{e}_x=(1,0,0)$, $\hvec{e}_y=(0,1,0)$ and $\hvec{e}_z=(0,0,1)$ uniformly discretized into a regular mesh with coordinate spacing $\Delta x$, $\Delta y$ and $\Delta z$.
Computational zones are centered at $(x_i,\, y_j,\, z_k)$ and delimited by the six interfaces orthogonal to the coordinate axis aligned, respectively, with $(x_{i\pm\HALF},\, y_j,\, z_k)$, $(x_i,\, y_{j\pm\HALF},\, z_k)$ and $(x_i,\, y_j,\, z_{k\pm\HALF})$.

\begin{figure}
  \centering
  \includegraphics[width=0.5\textwidth]{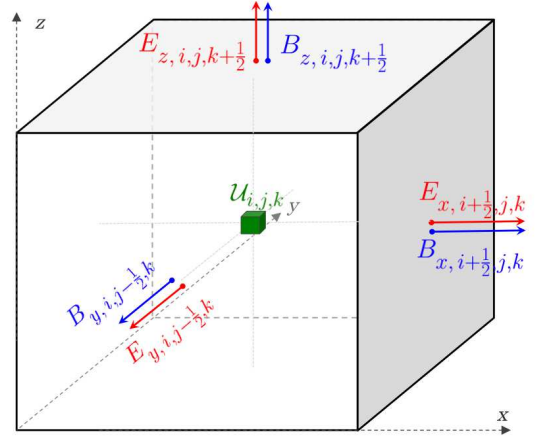}
  \caption{Positioning of hydrodynamical variable ($\cU_\cc$ in green) and
  electromagnetic fields ($\vec{E}_{\cf}$ and $\vec{B}_{\cf}$, red and blue respectively)
  inside a computational zone $(i,j,k)$.}
  \label{fig:ct}
\end{figure}

Primary zone-centered flow variables include density, momentum and energy and are stored by their volume averages inside the zone and labeled as $\cU_{\cc} = (D,\, \vec{m},\, \cE)_{\cc}$ where the $\cc$ subscript is a shorthand notation for $(i,j,k)$.

Conversely, electric and magnetic fields share the same staggered representation and are understood as surface-averaged quantities located at cell interfaces, as shown in Fig. \ref{fig:ct}.
We denote them as $\vec{E}_{\cf}, \vec{B}_{\cf}$ where the ${\cf}$ subscript tags the different face-centered electric and magnetic field components, e.g.,
\begin{equation}
  \vec{B}_{\cf} \equiv \left( \begin{array}{l}
              B_{\xf}  \\ \noalign{\medskip}
              B_{\yf}  \\ \noalign{\medskip}
              B_{\zf} \end{array}\right)
    = \left( \begin{array}{l}
              B_{x,i+\HALF,j,k}  \\ \noalign{\medskip}
              B_{y,i,j+\HALF,k}  \\ \noalign{\medskip}
              B_{z,i,j,k+\HALF} \end{array}\right) \,,
\end{equation}
and likewise for the face-centered electric field $\vec{E}_{\cf}$.
The subscripts $\xf$, $\yf$ and $\zf$ identify the different component as well as their staggered location inside the control volume, i.e., $\xf\equiv (x,i+\HALF,j,k)$, $\yf\equiv (y,i,j+\HALF,k)$ and $\zf\equiv (z,i,j,k+\HALF)$.
Within the CT approach, it also convenient to define the location of cell edges by introducing the ``e'' subscript, i.e., $\xe\equiv(x,i,j+\HALF,k+\HALF)$, $\ye\equiv(y,i+\HALF,j,k+\HALF)$ and $\ze\equiv(z,i+\HALF,j+\HALF,k)$.
Our formulation draws on the two fluid approach of \cite{Balsara_etal2016} and assumes a staggered representation for both the electric and magnetic field.
Zone-centered variables are updated using the standard finite volume approach.
Staggered quantities are updated using a discrete version of Stokes' theorem.

Choosing a staggered representation for $\vec{E}$ may seem unusual and somehow not consistent with the ideal limit, where $\vec{E}=-\vec{v}\times\vec{B}$ is not a continuous quantity across an interface.
However, this choice is consistent with a frozen Riemann solver where (in absence of source terms) the Ampere's law retains the same form as the induction equation thus leading, for a 1D problem, to the $E_x = const$ assumption.

In the following we will make frequent use of the backward difference operators $\Delta_x$, $\Delta_y$ and $\Delta_z$ defined as
\begin{equation}\label{eq:deltaOp}
  \begin{array}{l}
    \Delta_x Q_\cc \equiv  Q_\cc - Q_{\cc-\hvec{e}_x}
    \\ \noalign{\medskip}
    \Delta_y Q_\cc \equiv  Q_\cc - Q_{\cc-\hvec{e}_y}
    \\ \noalign{\medskip}
    \Delta_z Q_\cc \equiv  Q_\cc - Q_{\cc-\hvec{e}_z}  \,,
  \end{array}  
\end{equation}
where $Q$ can be any flow quantity.
The $\Delta$ operators can be equivalently applied to face-centered or edge-centered values.

In a similar fashion, we also introduce the face-to-center average operators which, to second-order, read
\begin{equation}\label{eq:avop}
  \begin{array}{l}
    \DS \av{Q_\xf}_x \equiv \frac{Q_\xf + Q_{\xf-\hvec{e}_x}}{2}  \\ \noalign{\medskip}
    \DS \av{Q_\yf}_y \equiv \frac{Q_\yf + Q_{\yf-\hvec{e}_y}}{2}  \\ \noalign{\medskip}
    \DS \av{Q_\zf}_z \equiv \frac{Q_\zf + Q_{\zf-\hvec{e}_z}}{2}  \,.
  \end{array}
\end{equation}

\subsection{IMEX Runge-Kutta Time Stepping}
\label{sec:imex}

A major challenge when dealing with the numerical solution of Eq. (\ref{eq:cons_law}) is the presence of a stiff source term in Ampere's law.
Owing to small resistivities typical of astrophysical plasma, this equation may easily become stiff and, for an almost ideal plasma (very small $\eta$), it cannot be solved through an explicit method in a computationally efficient way as the stiff time scale would impose prohibitively small time steps.
In order to overcome the time step restriction, we rely on the strong-stability-preserving (SSP) IMplicit-EXplicit (IMEX) Runge-Kutta method, introduced by \cite{Pareschi2005} and already employed in the context of the resistive RMHD equations by several authors, as cited in the introduction.
Applying the IMEX formalism to the original conservation law (\ref{eq:cons_law}), the resulting time-stepping scheme is explicit in $\cR(\cU)$ and implicit in the stiff source term $\cS(\cU)/\eta$.

We employ the $2^{\rm nd}$ order IMEX-SSP2(2,2,2) scheme which, when applied to the system (\ref{eq:cons_law}), consists of the following three stages,
\begin{equation}
  \begin{array}{l}
  \DS  \cU^{(1)} = \cU^n + a\frac{\Delta t}{\eta}\cS^{(1)}
  \\ \noalign{\medskip}
  \DS  \cU^{(2)} = \cU^{n} + \Delta t\cR^{(1)}
       + \frac{\Delta t}{\eta}\left[(1-2a)\cS^{(1)} + a\cS^{(2)}\right]
  \\ \noalign{\medskip}
  \DS  \cU^{n+1} = \cU^{n} + \frac{\Delta t}{2}
                  \left(\cR^{(1)} + \cR^{(2)}\right)
                 +\frac{\Delta t}{2\eta} \left[\cS^{(1)} + \cS^{(2)}\right]
 \end{array}
\end{equation}
where $a = 1-1/\sqrt{2}$ and the array $\cU = (D,\, \vec{m},\, \cE,\, \vec{B},\, \vec{E})$ contains zone-centered as well as face-centered conserved quantities.
The arrays $\cR(\cU)=-\nabla\cdot\tens{F}(\cU)+\cS_e(\vec{\cU})$ and $\cS(\cU)$ defined in the previous section embed, respectively, the explicit and implicit contributions.
It is important to realize that the first and second stage of the previous time-marching scheme are (locally) implicit as the source term is evaluated at the same intermediate stage as the left hand side.
This step involves therefore an implicit update of the electric field which, in our case, is located at zone faces.
This is discussed in Section \ref{sec:implicit_step}.

Without loss of generality, we write the single integration stage $s$ of a SSP Runge-Kutta-IMEX scheme by separately working out the zone-centered variables $\cU_\cc=(D,\, \vec{m},\, \cE)_\cc$ from the staggered fields $\vec{B}_\cf,\, \vec{E}_\cf$ as
%
%
\begin{align}
  \cU_\cc^{(s)} &=\DS  \cU_\cc^{n}
        - \Delta t\sum_{p=1}^{s-1} \tilde{a}_{sp}\Big(\nabla\cdot\tens{F}\Big)^{(p)}_\cc
    \label{eq:Uc_update} \\ \noalign{\medskip}
  \vec{B}_\cf^{(s)} &=\DS  \vec{B}_\cf^{n}
        - \Delta t\sum_{p=1}^{s-1}\tilde{a}_{sp}\Big(\nabla\times\vec{E}\Big)^{(p)}_\cf
    \label{eq:Bf_update} \\ \noalign{\medskip}
  \vec{E}_\cf^{(s)} &=\DS \vec{E}_\cf^{n}
       + \Delta t\left[\sum_{p=1}^{s-1}\tilde{a}_{sp}
          \left(\nabla\times\vec{B} - q\vec{v}\right)^{(p)}_\cf
        - \sum_{p=1}^{s}a_{sp}\tJ^{(p)}_\cf\right]  \label{eq:Ef_update}       
\end{align}
where $\tilde{a}_{sp}$ and $a_{sp}$ are the IMEX coefficients relative to the explicit and implicit temporal discretization \citep[see the Butcher tableau's available from tables II, III and IV in][]{Pareschi2005}, respectively.
The tensor $\tens{F}$ incorporates the flux components of the hydrodynamical variables only (columns 1-5 of Eq. \ref{eq:fluxTensor}).

At the implementation level, a generic IMEX stage can be decomposed into a sequence of explicit and implicit steps, that is, by first evolving Eqns. (\ref{eq:Uc_update})--(\ref{eq:Ef_update}) without the last source term in Ampere's law to some intermediate value and then by solving for the implicit source term alone.
The explicit and implicit steps are described in the next two sections.

\subsection{Recovery of Primitive Variables}
\label{sec:U2V}
%

Although the primary set to be evolved in time is that of conserved variables ($\cU_\cc$, $\vec{E}_\cf$ and $\vec{B}_\cf$), primitive variables defined by $\cV=(\rho,\vec{u},p,\vec{E},\vec{B})$ are required at any stage of the IMEX time stepping.
The conversion from primitive to conservative variables poses no difficulty, but the inverse transformation cannot be written in closed analytical form and must be recovered numerically.
Here we adopt the approach of \cite{DumbserZanotti2009} in which the problem is reduced to the solution of a quartic function in the Lorentz factor,
\begin{equation}
 \begin{array}{l}
  \DS   \left(C_1 - C_2^2\right)\gamma^4
   + 2C_2\frac{D}{\Gamma_1}\gamma^3
   \\ \noalign{\medskip}
  \DS + \left(C_2^2 - 2\frac{C_1}{\Gamma_1}-\frac{D^2}{\Gamma_1^2}\right)\gamma^2
       - 2\frac{C_2}{\Gamma_1}D\gamma
       + \frac{C_1 + D^2}{\Gamma^2_1} = 0 \,,
 \end{array}
\end{equation} 
where $C_1 = (\vec{m} - \vec{E}\times\vec{B})^2$ and $C_2 =  (\cE-\cP_{\rm EM})$ are, respectively, the square of the hydrodynamical momentum and the energy.
In the expressions above we have assumed an ideal equation of state,
\begin{equation}\label{eq:eos}
  w = \rho + \Gamma_1p \,,
\end{equation}
where $\Gamma_1 = \Gamma/(\Gamma - 1)$ and $\Gamma$ is the (constant) specific heat ratio.

As shown in the appendix of \cite{Zenitani_etal2009}, physically consistent solutions correspond to the larger of the two real roots of the previous quartic function.
We solve the quartic function using a combination of brackets, bisection and Newton-Raphson method.
Once the Lorentz factor is found, the remaining primitive variables can be computed as:
\begin{equation} \label{eq:recprimitive}
  \rho  = \frac{D}{\gamma}
  \,,\quad
  p     = \frac{\cE - \rho\gamma^2 - \cP_{\rm EM}} 
               {\Gamma_1\gamma^2 - 1}
  \,,\quad
  \vec{v}  = \frac{\vec{m} - \vec{E}\times\vec{B}}{(\rho + \Gamma_1 p)\gamma^2}\,,
\end{equation}
where the second one is obtained from the energy equation (\ref{eq:totE}) while $\cP_{\rm EM}$ - the EM energy density - has been defined after Eq. (\ref{eq:totE}).

A distinct inversion scheme \citep[used, for instance, by][]{Dionys_etal2013} consists of subtracting the electromagnetic contributions from momentum and energy densities and then resorting to a standard relativistic hydro inversion scheme to find the pressure.
This can easily achieved using, e.g., the approach outlined by \cite{MPB2005} which is also valid for different equations of state.

\subsection{Explicit Step}
\label{sec:explicit_step}
%

We now describe the spatial discretization adopted for the evolution of the zone-centered and staggered variables during the explicit stage of an IMEX stage (Eqns \ref{eq:Uc_update}-\ref{eq:Ef_update} without the last term in Ampere's law).

\subsubsection{Explicit Update of Zone-Centered Quantities}
\label{sec:center_update}
%
In the finite volume approach, Eq. (\ref{eq:Uc_update}) is naturally interpreted as an integral relation relating the change of a volume-averaged conserved quantity to its surface-averaged flux integral across the cell boundary.
The discrete form of the divergence operator is thus computed using fluid and electromagnetic quantities available at the $p$-th stage of the time-marching scheme: 
\begin{equation}\label{eq:divH}
 \left(\nabla\cdot\tens{F}\right)^{(p)}_\cc
  =   \frac{\Delta_x\cF^*_\xf}{\Delta x}
    + \frac{\Delta_y\cF^*_\yf}{\Delta y}
    + \frac{\Delta_z\cF^*_\zf}{\Delta z}  \,,
\end{equation}
where the $\Delta$'s are the backward difference operators defined in Eq. (\ref{eq:deltaOp}).
Here the different $\cF^{*}$ are the hydrodynamic components of the flux computed with a one-dimensional Riemann solver applied at cell interfaces between left and right states at the $(p)$-th stage.
To second-order accuracy, a midpoint quadrature rule suffices so that, e.g., 
\begin{equation}
   \cF^*_\xf =
   \cF_{\rm Riem} \Big(\cV^L_\xf,\, \cV^R_\xf \Big) \,,
\end{equation}
where $\cV^L_\xf$ and $\cV^R_\xf$ are the one-sided limit values of the piecewise polynomial reconstruction from within the two neighbor zones adjacent to the interface \citep[higher than second-order finite-volume schemes requires more quadrature points, see, for instance, in][]{Balsara_etal2016}.
The reconstruction is carried out on the set of primitive variables $\cV$ (rather than $\cU$) as it is known to produce less oscillatory results.
For a second-order reconstruction one has, e.g., 
\begin{equation}
  \cV^L_\xf = \cV_\cc              + \frac{\delta_x \cV_\cc}{2} \,,\quad
  \cV^R_\xf = \cV_{\cc+\hvec{e}_x} - \frac{\delta_x \cV_{\cc + \hvec{e}_x}}{2} \,,
\end{equation}
where $\delta_x\cV_\cc$ are limited slopes in the $x$-direction.
A widespread choice, which will also be used by default in the present work, is the van Leer (or harmonic mean) limiter:
\begin{equation}\label{eq:VL_lim}
  \delta_x\cV_{\cc} = \left\{\begin{array}{ll}
  0   & {\rm if} \;
        (\Delta_x\cV_\cc)(\Delta_x\cV_{\cc+\hvec{e}_x})  < 0
  \\ \noalign{\medskip}
  \DS \frac{2 (\Delta_x\cV_\cc) (\Delta_x\cV_{\cc+\hvec{e}_x})}
       {  (\Delta_x\cV_\cc) + (\Delta_x\cV_{\cc+\hvec{e}_x})}\
      & {\rm otherwise}\,.
  \end{array}\right.
\end{equation}
A slightly more compressive option is given by
\begin{equation}\label{eq:MC_lim}
  \delta_x\cV_{\cc} = {\rm mm}\left(
    \frac{\Delta_x\cV_{\cc} + \Delta_x\cV_{\cc+\hvec{e}_x}}{2},\,
    2{\rm mm}(\Delta_x\cV_{\cc},\Delta_x\cV_{\cc+\hvec{e}_x})\right)
\end{equation}
where $\rm{mm}(\cdot,\cdot)$ is the minmod limiter.
Eq. (\ref{eq:MC_lim}) is known as the monotonized central (MC) limiter.

A popular choice for solving the Riemann problem at cell interfaces in the context of Res-RMHD is the Lax-Friedrichs (LF) scheme, where the fastest characteristic speed is chosen to be the speed of light,
\begin{equation}\label{eq:LF_Solver}
  \cF^*_{\xf} =   \frac{\tens{F}_x(\cV^R_\xf) + \tens{F}_x(\cV^L_\xf)}{2}
                - \frac{\cU^R_{\xf} - \cU^L_{\xf}}{2} \,.
\end{equation}
One-dimensional flux functions in the $y$- and $z$-directions are computed in a similar fashion.
Despite its simplicity, the LF scheme (\ref{eq:LF_Solver}) can lead to excessive smearing of discontinuous waves as only the two outermost (light) waves are retained in the Riemann fan.
For this reason we derive, in section \ref{sec:method_riemann}, an improved five-wave solver designed to capture the light waves as well as the intermediate acoustic modes, including the contact discontinuity.
Our approach is similar to the Harten-Lax-van Leer contact wave (HLLC) Riemann solver recently proposed in the appendix of \cite{Miranda_etal2018}.

\subsubsection{Explicit Update of Face-Centered Quantities}
\label{sec:face_update}
%

In the constrained-transport method, Eq. (\ref{eq:Bf_update}) and (\ref{eq:Ef_update}) are understood as surface-averaged relations so that a discrete version of Stokes' theorem can be applied.
This leads to the appearance of edge-centered values (or line-averaged integrals) of the electric and magnetic fields, that is, $\vec{E}_e\equiv(E_\xe, E_\ye, E_\ze)$ and $\vec{B}_e\equiv(B_\xe, B_\ye, B_\ze)$.
Exact integration over the control volume surfaces yields
\begin{equation}\label{eq:Bf_discrete}
  \begin{array}{r}
    \Big(\nabla\times\vec{E}\Big)^{(p)}_\cf =\DS
        \DS \left(  \frac{\Delta_yE^*_\ze}{\Delta y}
                  - \frac{\Delta_zE^*_\ye}{\Delta z}\right)_\xf\hvec{e}_x
  \\ \noalign{\medskip}
   \DS + \DS \left(  \frac{\Delta_zE^*_\xe}{\Delta z}
                  - \frac{\Delta_xE^*_\ze}{\Delta x}\right)_\yf\hvec{e}_y
  \\ \noalign{\medskip}
   \DS + \DS \left(  \frac{\Delta_xE^*_\ye}{\Delta x}
                  - \frac{\Delta_yE^*_\xe}{\Delta y}\right)_\zf\hvec{e}_z
  \end{array}                              
\end{equation}
while, for the Ampere's law we have similarly:
\begin{equation}\label{eq:Ef_discrete}
  \begin{array}{r}
    \Big(\nabla\times\vec{B} - q\vec{v}\Big)^{(p)}_\cf =
        \DS \left(  \frac{\Delta_yB^*_\ze}{\Delta y}
                  - \frac{\Delta_zB^*_\ye}{\Delta z} - (qv_x)^*\right)_\xf\hvec{e}_x
  \\ \noalign{\medskip}
   + \DS \left(   \frac{\Delta_zB^*_\xe}{\Delta z}
                - \frac{\Delta_xB^*_\ze}{\Delta x} - (qv_y)^*\right)_\yf\hvec{e}_y
  \\ \noalign{\medskip}
   + \DS \left(   \frac{\Delta_xB^*_\ye}{\Delta x}
                - \frac{\Delta_yB^*_\xe}{\Delta y} - (qv_z)^*\right)_\zf\hvec{e}_z
  \end{array}                              
\end{equation}
The $\Delta$ operators are again defined in Eq. (\ref{eq:deltaOp}).

The edge-centered electric and magnetic fields are tagged with a star and are obtained by  using a two-dimensional Riemann solver at the $(p)$-th stage of integration.
These upwind flux functions result indeed from a four-state representation since two surfaces of discontinuity intersect at a zone edge so that modes of Riemann fans coming from different directions overlap \citep[see, for instance,][]{dZBL2003, LdZ2004, Balsara_etal2016}.
Since the Maxwell's equations are linear and involve propagation of light waves (in the frozen limit approach used here), a proper combination of upwind numerical fluxes along the corresponding orthogonal coordinates leads to the single-valued numerical flux out of these four-states.
For the $z$-components one has, for instance,
\begin{equation}\label{eq:maxwell_solver_Ez}
  \begin{array}{ll}
  E^*_\ze =& \DS   \frac{1}{4}\left[  E^{LL}_\ze + E^{LR}_\ze
                                    + E^{RL}_\ze + E^{RR}_\ze\right]
  \\ \noalign{\medskip}                                   
   &\DS  - \frac{1}{2} (B^R_{\xf+\HALF\hvec{e}_y} - B^L_{\xf+\HALF\hvec{e}_y})
  \\ \noalign{\medskip}                                   
   &\DS  + \frac{1}{2} (B^R_{\yf+\HALF\hvec{e}_x} - B^L_{\yf+\HALF\hvec{e}_x})\,,
  \end{array}                
\end{equation}
and likewise for the $z$-component of the edge-centered magnetic field,
\begin{equation}\label{eq:maxwell_solver_Bz}
  \begin{array}{ll}
  B^*_\ze =& \DS   \frac{1}{4}\left[  B^{LL}_\ze + B^{LR}_\ze
                                    + B^{RL}_\ze + B^{RR}_\ze\right]
  \\ \noalign{\medskip}                                   
   &\DS + \frac{1}{2} (E^R_{\xf+\HALF\hvec{e}_y} - E^L_{\xf+\HALF\hvec{e}_y})
  \\ \noalign{\medskip}                                   
   &\DS - \frac{1}{2} (E^R_{\yf+\HALF\hvec{e}_x} - E^L_{\yf+\HALF\hvec{e}_x})\,. 
  \end{array}                
\end{equation}

\begin{figure}
  \centering
  \includegraphics[width=0.5\textwidth]{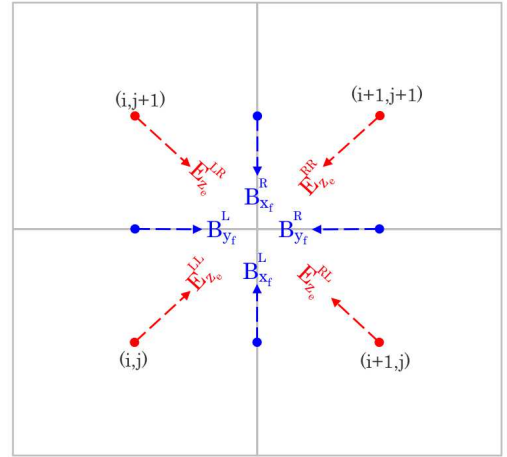}
  \caption{Top view of the four zones abutting the $z$-edge (intersection of squares)
  showing a schematic representation of the reconstruction steps needed to compute
  the line-averaged $z$-component of the electric field in the 2D Riemann solver.
  The third dimension has been collapsed for the sake of clarity.
  Red and blue arrows indicate, respectively, the direction of the reconstruction
  for the electric and magnetic field, starting from the primary location of
  the variable (represented by a dot).}
  \label{fig:maxwell_solver}
\end{figure}
The four terms with a double superscript in square bracket denote reconstructed values of the $z$-components of the electric and magnetic fields from the face centers (where they are primarily defined) to the zone edge $\ze$.
This is schematically illustrated in Fig. \ref{fig:maxwell_solver} where a top view of the four zones intersecting at an edge (represented by the central point common to all squares) is shown.
The first and second superscript refer to the state lying to the left (L) or to the right (R) of the edge with respect to the $x$- and $y$-direction, respectively.
For a second-order reconstruction one may use, e.g., 
\begin{equation}
  \begin{array}{l}
  \DS E^{LL}_\ze =
      \av{E_\zf}_z + \av{\frac{\delta_x E_\zf}{2}}_z
                   + \av{\frac{\delta_y E_\zf}{2}}_z
  \\ \noalign{\medskip}
  \DS E^{RL}_\ze =
      \av{E_{\zf+\hvec{e}_x}}_z - \av{\frac{\delta_x E_{\zf+\hvec{e}_x}}{2}}_z
                                + \av{\frac{\delta_y E_{\zf+\hvec{e}_x}}{2}}_z
  \end{array}
\end{equation}
where the $\av{\cdot}_z$ is the arithmetic average operator in the $z$-direction, see Eq. (\ref{eq:avop}).

The $x$- and $y$-components of the staggered magnetic field, on the other hand, are continuous at $x$- and $y$-faces and the dissipative terms in Eq. (\ref{eq:maxwell_solver_Ez}) come from two independent wave fans involving jumps in the transverse directions.
The same arguments hold for the $x$- and $y$-components of the staggered electric field when evaluating the dissipative terms in Eq. (\ref{eq:maxwell_solver_Bz}).
Flux functions located at $x$- and $y$-edges are obtained similarly by cyclic index permutation.

The last term in square brackets in Eq. (\ref{eq:Ef_discrete}) is a non-stiff source responsible for passive charge advection and it is discretized, for reasons that will be clear in the following section, using an upwind selection rule.
We have explored two different options, giving essentially very similar results.
In the first one, slightly more diffusive, a simple Rusanov Lax-Friedrichs solver may be employed:
\begin{equation}\label{eq:qv1}
  (qv_x)_\xf^* =   \frac{(qv_x)^R_\xf + (qv_x)^L_\xf}{2}
                 - \frac{q^R_\xf - q^L_\xf}{2} \,,
\end{equation}
and similarly for at the $y$- and $z$-interfaces.
Alternatively, one could select the upwind direction by looking at the sign of the density flux $F^{*,[\rho]}_{\xf}$, computed during the evolution of zone-centered variables (Section \ref{sec:center_update}):
\begin{equation}\label{eq:qv2}
  (qv_x)_f^* = \left\{\begin{array}{ll}
    \DS \left(\frac{q}{D}\right)^L_\xf F^{*,[\rho]}_\xf
        & \quad {\rm if} \quad F^{*,[\rho]}_\xf > 0  \\ \noalign{\bigskip}
    \DS \left(\frac{q}{D}\right)^R_{x_f} F^{*,[\rho]}_\xf
        & \quad {\rm if} \quad F^{*,[\rho]}_\xf < 0   \,,
  \end{array}\right.
\end{equation}
where $D=\gamma\rho$ is the laboratory density.

The choice of an upwind rule ensures that the charge, computed as the discrete divergence of the electric field, remains oscillation-free (see the discussion at the end of Section \ref{sec:implicit_step} and \ref{sec:charge_conservation}).

\subsection{Implicit Step}
\label{sec:implicit_step}
%

During the implicit step one has to deal with
\begin{equation}\label{eq:implicitE}
  \vec{E}^{(s)} =  \vec{R}_e^{(s-1)} - \frac{\delta t}{\eta}
                   \Big[\gamma\vec{E} + \vec{u}\times\vec{B}
                                     - (\vec{E}\cdot\vec{u})\vec{v}\Big]^{(s)} \,,
\end{equation}
where $\delta t = a_{ss}\Delta t$, $\vec{R}^{(s-1)}_e$ contains only explicit terms, $\vec{B}^{(s)}$ is also computed explicitly, while the electric field $\vec{E}$ and four-velocity $\vec{u} = \gamma\vec{v}$ must be determined.
Eq. (\ref{eq:implicitE}) holds at any location inside the computational zone and has to be solved iteratively by adjusting primitive variables ($\rho$, $\vec{v}$ and $p$) while leaving laboratory density, momentum, energy and magnetic field unchanged.
In the proposed CT formalism, a major complication now arises since the primary electric field variable is located at zone faces while conserved variables retain a zone-centered representation.
An implicit update would have now to be carried at the three different zone faces, thus requiring simultaneous reconstructions from neighboring cells with the unwanted effect of making the algorithm non-local anymore.

This complication could be overcome if one notices that the implicit relation (\ref{eq:implicitE}) is linear in the electric field and could be easily inverted if the velocity field $\vec{u}$ at the next level is known.
This suggests that one could first solve the implicit step using the cell-centered electric field and then reconstruct the resulting velocity field at the faces.
Then Eq. (\ref{eq:implicitE}) could be readily solved for the staggered electric field yielding
\begin{equation}\label{eq:implicitE_solve}
  \vec{E} = \frac{\eta \vec{R}^{(s-1)}_e - \delta t
  \left[\vec{u}\times\vec{B} - (\vec{E}\cdot\vec{u})\vec{v}\right]}
                     {\eta + \gamma \delta t} \,,
\end{equation}
where now the velocity has been reconstructed using the values obtained during the cell-center implicit step and we have dropped the superscript $(s)$ for ease of notations.
The $(\vec{E}\cdot\vec{u})$ term (which contains also the transverse components of $\vec{E}$) can be rewritten by taking the scalar product of Eq. (\ref{eq:implicitE}) with $\vec{u}$:
\begin{equation}\label{eq:Eu}
   (\vec{E}\cdot\vec{u}) = \gamma\eta\frac{\vec{R}_e^{(s-1)}\cdot\vec{u}}
                                           {\gamma\eta + \delta t}  \,.
\end{equation}
Equation (\ref{eq:implicitE_solve}) together with (\ref{eq:Eu}) now directly gives the staggered components of $\vec{E}$ at the chosen interface once the values of velocity and magnetic fields are available at the same location.
The interface values of $\vec{u}$ and $\vec{B}$ can be obtained by interpolating the cell-centered four-velocity and staggered magnetic fields at the same location.
However, both $\vec{u}$ and the transverse components of $\vec{B}$ will, in general, be discontinuous at a given interface.

In order to overcome this problem, we have examined two different options.
In the first one, a single-value of velocity and magnetic field can be simply obtained by taking, e.g., the arithmetic average of the left and right states adjoining the interface.
Then Eq. (\ref{eq:implicitE_solve}) can be readily solved for the staggered electric field.
In the second option, Eq. (\ref{eq:implicitE_solve}) could be solved separately for the left and right states thus yielding two different values of the staggered electric field, $\vec{E}^L_f$ and $\vec{E}^R_f$.
A single-valued electric field can then be obtained by some averaging procedure and, although in principle several options are possible to produce a single-valued electric field, the rationale for choosing one average versus another is better understood by inspecting at the resulting discretization of the charge (Section \ref{sec:charge_conservation}).
Here we choose the simple arithmetic average
\begin{equation}\label{eq:arithm_averageE}
  \vec{E}_\cf = \frac{\vec{E}^L_\cf + \vec{E}^R_\cf}{2} \,,
\end{equation}
which from Eq. (\ref{eq:implicitE}) allows the (stiff part of the) current to be expressed as 
\begin{equation}\label{eq:arithm_averageJ}
  \tilde{\vec{J}}_\cf = \frac{\tJ^L_\cf + \tJ^R_\cf}{2} \,.
\end{equation}
As it will be later shown in Section \ref{sec:charge_conservation}, this choice is equivalent to a conservative update of the charge in which the interface fluxes are computed by means of a local Rusanov Lax-Friedrichs Riemann solver.
The amount of dissipation is introduced by the upwind discretization of the non-stiff source term $q\vec{v}$ (see the discussion at the end of Section \ref{sec:face_update}).
Although both options have been implemented in the code and no substantial difference has been found, we use the second options throughout this work.

The implicit solver for the cell-centered electric field is the subject of next sub-section.

%
%

\subsubsection{A Newton-Broyden Scheme for Cell-Center Implicit Update}
\label{sec:IMEX_NB}
%

We propose a novel method to solve Eq. (\ref{eq:implicitE}) at the cell center based on a multidimensional Newton-Broyden root-finder scheme.
Our method exploits conservation of momentum, energy and density through the iterative cycle so that, instead of solving directly Eq. (\ref{eq:implicitE}), we search for the roots of 
\begin{equation}\label{eq:NB}
 \vec{f}(\vec{u}) \equiv
 \vec{m} - \Big[Dh(\vec{u})\vec{u} + \vec{E}(\vec{u})\times\vec{B}\Big] = 0 \,,
\end{equation}
where the specific enthalpy $h = w/\rho$ and the electric field are functions of the four-velocity while $\vec{m}\equiv\vec{m}^{(s)}$, $D\equiv D^{(s)}$ and $\vec{B}\equiv\vec{B}^{(s)}$ are known quantities at the beginning of the implicit step.
Equation (\ref{eq:NB}) gives a nonlinear system of equations in the four-velocity since $\vec{E}=\vec{E}(\vec{u})$ through Eq. (\ref{eq:implicitE_solve}) while $h=h(\vec{u})$ through the energy equation (\ref{eq:totE}).
It simply states that the momentum must not change during the implicit update.
We use the four velocity (rather than the electric field) as the independent variable as it offers the advantage that conservative to primitive inversion can be avoided during the cycle.

The Newton-Brodyen method can be sketched as follows:
\begin{enumerate}
\setlength\itemsep{1.2em}

\item
At the beginning of the step, compute the electric field at zone centers by simple arithmetic average of the face value,
\begin{equation}
  \hvec{e}_d\cdot\vec{E}_0 = \av{\hvec{e}_d\cdot\vec{E}}_{d} \,,
\end{equation}
where $d=x,y,z$.
Also, provide a suitable guess of the four-velocity\footnote{
We employ the four velocity $\vec{u}_0$ at the current time level if $\eta/\Delta t > 1$; otherwise we use the four-velocity obtained from the ideal RMHD equations}.
The iteration counter is set to $\kappa = 0$;

\item
\label{item:NB_begcycle}
Using the current value of $\vec{u}^{(\kappa)}$, obtain the improved values of the cell-centered electric field using Eq. (\ref{eq:implicitE_solve}) as well as pressure and enthalpy using Eq. (\ref{eq:recprimitive});

\item
Compute $\vec{f}^{(\kappa)} \equiv \vec{f}(\vec{u}^{(\kappa)})$:
\begin{equation}\label{eq:NB_solve}
  \vec{f}^{(\kappa)} = \vec{m} - \Big[Dh(\vec{u})\vec{u}
                               + \vec{E}(\vec{u})\times\vec{B}\Big]^{(\kappa)} \,.
\end{equation}
Note that the laboratory density, magnetic field and total energy cannot change during this step.

\item
Using the Jacobian, $\tens{J}=\partial\vec{f}/\partial\vec{u}$, obtain an improved guess of the four-velocity through
\begin{equation}
 \vec{u}^{(\kappa+1)} =  \vec{u}^{(\kappa)}
                        - \left(\tens{J}^{(\kappa)}\right)^{-1}\vec{f}^{(\kappa)}\,,
\end{equation}
where
\begin{equation}\label{eq:f_NB}
  \tens{J}_{ij} = - Dh(\vec{u})\delta_{ij}
                  - Du_i\pd{h(\vec{u})}{u_j}
                  - \pd{(\vec{E}(\vec{u})\times\vec{B})_i}{u_j} \,,
\end{equation}
and an explicit derivation is presented in Appendix \ref{app:NB_Jacobian}.

\item
Exit from the iteration cycle if the error is less than some prescribed accuracy $|\vec{f}^{(\kappa)}|<\epsilon$, otherwise we go back to step \ref{item:NB_begcycle} and let $\kappa\to \kappa+1$.

\end{enumerate}

Note that conversion from conservative to primitive variables is \emph{not} required at each cycle since the primitive variables are automatically updated.
For standard applications, our algorithm converges in $2-5$ iterations with a relative tolerance of $10^{-11}$.

\subsection{Charge Conservation}
\label{sec:charge_conservation}
%

In the CT approach, the charge is not an independent variable (as it is the case for the GLM), but it is directly obtained from the divergence of the electric field,
\begin{equation}\label{eq:charge}
  q_\cc \equiv (\nabla\cdot\vec{E})_c
        =      \frac{\Delta_xE_\xf}{\Delta x}
             + \frac{\Delta_yE_\yf}{\Delta y}
             + \frac{\Delta_zE_\zf}{\Delta z} \,.
\end{equation}
The charge density is thus collocated at the cell center and it is conserved by construction.
In order to see this, let us take the discrete divergence of Ampere's law (Eq. \ref{eq:Ef_update}) at any integration stage.
Since CT schemes automatically fulfill the condition $\nabla\cdot\nabla\times = 0$ at the discrete level, one finds
\begin{equation}
   q^{(s)}_\cc = q^{n}_\cc - \Delta t\nabla\cdot\left(
     \sum_{p=1}^{s-1}\tilde{a}_{sp}(q\vec{v}^*)^{(s)}
   + \sum_{p=1}^s            a_{sp}\tJ^{(s)}\right)_\cc \,.
\end{equation}
The previous relation shows that the charge density obeys a conservative equation by construction.

We now show that when the choices given by Eq. (\ref{eq:arithm_averageJ}) and Eq. (\ref{eq:qv2}) are applied to Ampere's law, one obtains a stable and oscillation-free equivalent discretization for the charge evolution.
For simplicity we restrict our attention to the $1^{\rm st}$ order IMEX scheme which, using our standard notations, reads
\begin{equation}
  \begin{array}{lcl}
  \vec{E}^{(1)}_\cf  & = & \vec{E}^n_\cf - \Delta t\tJ^{(1)}_\cf
  \\ \noalign{\medskip}
  \vec{E}^{n+1}_\cf  & = &\DS  \vec{E}^n_\cf + \Delta t(\nabla\times\vec{B}^n)_\cf
       - \Delta t\left[\tJ^{(1)}_\cf + (q\vec{v})^*_\cf\right]
  \end{array}
\end{equation}
By taking the divergence of the last expression and using Eq. (\ref{eq:arithm_averageJ}) and Eq. (\ref{eq:qv2}) one finds
\begin{equation}
  q^{n+1}_\cc = q^n_\cc - \Delta t\left(
                     \frac{\Delta_xJ^*_\xf}{\Delta x}
                   + \frac{\Delta_yJ^*_\yf}{\Delta y}
                   + \frac{\Delta_zJ^*_\zf}{\Delta z} \right)
\end{equation}
and using Eq. (\ref{eq:qv1}), for instance, one finds that
\begin{equation}
  \begin{array}{lcl}
  J^*_\xf &=&\DS \frac{   \left(\tilde{J}_\xf^{(1)} + (qv_\xf)^{n}\right)^L
                         +\left(\tilde{J}_\xf^{(1)} + (qv_\xf)^{n}\right)^R }
                       {2}
  \\ \noalign{\medskip}                                 
          & & \DS - \frac{q^{n,R}_\xf - q^{n,L}_\xf}{2}
  \end{array}          
\end{equation}
is a Rusanov Lax-Friedrichs numerical flux.

\subsection{Summary of the CT Method}
%

Hereafter we summarize the main steps followed by our IMEX SSP2-CT scheme in order to ease up the implementation of the different computational tasks.

\begin{enumerate}
\setlength\itemsep{1.2em}

\item
At the beginning of integration, we start with $\cU^{n}_\cc$, $\vec{E}^n_\cf$ and $\vec{B}^n_\cf$ as our primary fields.
Set $s=1$.

\item
\label{item:CT:implicit_c}
Average the primary staggered fields ($\vec{E}$ and $\vec{B}$) to zone centers, using Eq. (\ref{eq:avop}).
Perform the IMEX implicit update at zone centers to obtain $\vec{E}^{(s)}_\cc$ and primitive hydrodynamic variables $\cV^{(s)}$, as described in Section \ref{sec:IMEX_NB}.

\item
\label{item:CT:implicit_s}
Using $\vec{E}^{(s)}_\cc$ and $\vec{u}^{(s)}_\cc$, achieve the implicit step on the staggered electric field as well, as detailed in Section \ref{sec:implicit_step}.
This yields $\vec{E}^{(s)}_\cf$.

\item
\label{item:CT:explicit}
Compute interface Godunov fluxes and the multidimensional line-averaged electric and magnetic fields to obtain the divergence and curl operators contained in the predictor step ($s=2$) of the IMEX scheme, that is, Eqns (\ref{eq:Uc_update}, \ref{eq:Bf_update}, \ref{eq:Ef_update}) for $s=2$.
Also, compute the explicit source terms needed for this stage and add all terms up to obtain the explicit contributions needed for $\cU^{(2)}_\cc$, $\vec{B}^{(2)}_\cf$ and $\vec{E}^{(2)}_\cf$. 

\item
Repeat the implicit steps \ref{item:CT:implicit_c} and \ref{item:CT:implicit_s} with $s=2$ to achieve $\cU^{(2)}_\cc$ $\vec{E}^{(2)}_\cf$.

\item Perform the final corrector stage as in step \ref{item:CT:explicit} and obtain the solution at the next time level $n+1$.

\end{enumerate}

\section{The MHLLC Riemann Solver}
\label{sec:method_riemann}
%
%
%

Within the IMEX formalism, the solution to the Riemann problem can be obtained under the frozen limit condition (infinite resistivity) by ignoring the effect of the stiff source term in Ampere's law on the characteristic wave propagation.
In this limit, the current-density is $\vec{J}=q\vec{v}$ and the momentum and energy equations in (\ref{eq:ResRMHD}) can be rewritten as
\begin{equation}\label{eq:ResRMHD_inf_eta}
  \arraycolsep=4pt
  \begin{array}{lcl}
    \DS  \pd{}{t}(w\gamma^2\vec{v})
    + \nabla\cdot\Big(w\gamma^2\vec{v}\vec{v} + \tens{I}p\Big) &=&
      q(\vec{E} + \vec{v}\times\vec{B})
    \\ \noalign{\medskip}
    \DS \pd{}{t}(w\gamma^2 - p) + \nabla\cdot(w\gamma^2\vec{v}) &=& 
     q\vec{v}\cdot\vec{E} \,.
  \end{array}  
\end{equation}
Thus the coupling between EM and hydrodynamic fields results only from a non-zero charge density.
At the linear level \cite{MMB2018} have shown that, in this regime, the characteristic structure of the resistive RMHD equations entails a pair of acoustic modes and a pair of light waves.
The contribution of the charge, being a second-order (nonlinear) term, is neglected in the dispersion relation.
In addition, assuming that the presence of the source term on the right hand side of Eq. (\ref{eq:ResRMHD_inf_eta}) does not change the Rankine-Hugoniot conditions \citep[a similar inference has been done by][]{Miranda_etal2018}, we can consider Maxwell's and hydrodynamic equations to be decoupled during the solution of the Riemann problem.

\begin{figure}
  \centering
  {\includegraphics[trim={1.5cm 4cm 4.0cm 7cm}, width=0.45\textwidth]{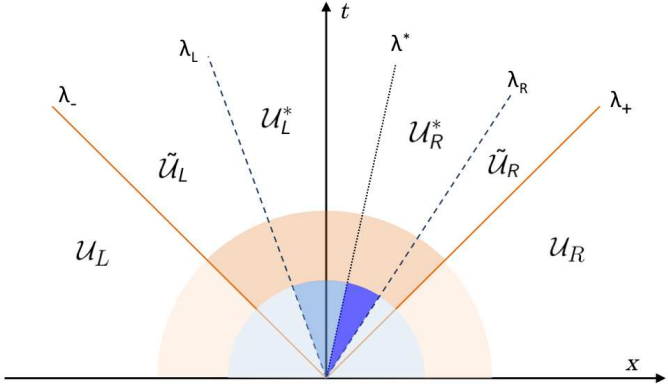}}
  \caption{Schematic representation of the Riemann fan assumed by the MHLLC solver.
           The outermost waves ($\lambda_{+} = -\lambda_{-} = 1$) are EM (light) waves
           while $\lambda_L$ and $\lambda_R$ are pairs of acoustic waves.
           The central mode ($\lambda^*$) represents a contact discontinuity.
           Jumps in EM fields are denoted by different shades of orange while jumps
           of hydrodynamic variables are marked by different shades of blue.}
  \label{fig:MHLLC}
\end{figure}
Our approach therefore is based on the direct combination of two Riemann solvers: one for the outermost EM waves across which only transverse components of electric and magnetic fields can change, and a second one across the sound waves where only hydrodynamical variables have non-trivial jumps.
We shall refer to the former and the latter as the \emph{outer} and the \emph{inner} Riemann solver, respectively.
Across the outermost EM waves, jump conditions for $\vec{E}$ and $\vec{B}$ follow directly from Maxwell's equations.
The presence of the outermost waves modifies total momentum and energy so that conservative variables behind the discontinuities serve as input left and right states to the inner Riemann problem for which any any relativistic hydrodynamic solver may be employed.
Here we choose to rely on relativistic HLLC scheme of \cite{MB2005}.

Across each wave $\lambda$ the jump conditions
\begin{equation}\label{eq:rankine}
   \lambda (\cU_r - \cU_l) = F_r - F_l
\end{equation}
must be satisfied for \emph{any} pair of states $\cU_r$ and $\cU_l$ and corresponding flux functions $F_l$ and $F_r$.
The Riemann fan comprises five waves and its structure is schematically depicted in Fig. \ref{fig:MHLLC}.
For a discontinuity propagating in the $n$ ($=1,2,3$) direction, our state and flux functions are more conveniently written by separating hydrodynamical from EM terms as
\begin{equation}
  \cU = \left(\begin{array}{c}
    D         \\ \noalign{\medskip}
    Q_i + S_i \\ \noalign{\medskip}
    \cE_h + \cP_{\rm EM}  \\ \noalign{\medskip}
    B_i       \\ \noalign{\medskip}
    E_i
   \end{array}\right)
   ,\quad
   \cF = \left(\begin{array}{c}
        Dv_n                              \\ \noalign{\medskip}
        Q_iv_n + p\delta_{in} + \tens{T}_{in} \\ \noalign{\medskip}
        Q_n + S_n                       \\ \noalign{\medskip}
         \epsilon_{ink}E_k                 \\ \noalign{\medskip}
        -\epsilon_{ink}B_k
   \end{array}\right)
\end{equation}
where $\tens{T}$ is the Maxwell stress tensor (Eq. \ref{eq:MaxwellStress}) while 
\begin{equation}
  Q_i   = w\gamma^2 v_i   \,,\quad
  \cE_h = w\gamma^2 - p   \,,\quad
  S_i   = (\vec{E}\times\vec{B})_i 
\end{equation}
are the hydrodynamic momentum, energy and the EM Poynting vector, respectively.

Owing to its structure, we label our Riemann solver as \quotes{MHLLC}, where \quotes{M} denotes the outer Maxwell solver while \quotes{HLLC} stands for the inner the Harten-Lax-van Leer-Contact approximate Riemann solver originally introduced by \cite{TSS1994} in the context of the classical equations of gas-dynamics.
The HLLC scheme improves over the traditional HLL method \citep{HLL1983,ERMS1991} by restoring the contact wave in the solution.
We point out that, although devised from different assumptions, our method of solution ends up being similar to the version of the HLLC solver reported in the appendix of \cite{Miranda_etal2018}.
The outer and inner solvers are described in the next two sections.

\subsection{Outer Solver}
%

The jump conditions (\ref{eq:rankine}) must hold across the leftmost wave ($\lambda_-=-1$) between states $\cU_L$ and $\tilde{\cU}_L$ and, likewise, across the rightmost wave ($\lambda_+=1$) between $\cU_R$ and $\tilde{\cU}_R$.
In the frozen limit, EM fields can be discontinuous across the outermost waves but do not experience further jumps across the inner waves $\lambda_L$ and $\lambda_R$.
For this reason we set $\tilde{\vec{E}}_L = \vec{E}^*_L=\vec{E}^*_R = \tilde{\vec{E}}_R$ (and similarly for $\vec{B}$).
Specializing to the $x$-direction and solving Maxwell's jump conditions for the transverse components yields
\begin{equation}\label{eq:resHLLC_EBtilde}
  \begin{array}{lcl}
    \tilde{B}_y &=&\DS \frac{B_{y,L} + B_{y,R}}{2} + \frac{E_{z,R} - E_{z,L}}{2}
  \\ \noalign{\medskip}
    \tilde{B}_z &=&\DS \frac{B_{z,L} + B_{z,R}}{2} - \frac{E_{y,R} - E_{y,L}}{2}
  \\ \noalign{\medskip}
    \tilde{E}_y &=&\DS \frac{E_{y,L} + E_{y,R}}{2} - \frac{B_{z,R} - B_{z,L}}{2}
  \\ \noalign{\medskip}
    \tilde{E}_z &=&\DS \frac{E_{z,L} + E_{z,R}}{2} + \frac{B_{y,R} - B_{y,L}}{2} \,.
  \end{array}
\end{equation}
Normal components $B_x$ and $E_x$ are continuous at the interface and do not experience any jump.

With the above definitions, the continuity equation is trivially fulfilled since both density and fluid velocity are continuous across $\lambda_\pm$.
Likewise, it can also be verified that the jump conditions (\ref{eq:rankine}) of the momentum and energy equations are also automatically satisfied.
In particular, by combining the jump conditions of the $x$-component of the momentum and energy, one consistently finds
\begin{equation}\label{eq:inner_mx+E}
  \begin{array}{lcl}
    (S_{xL} - \tilde{S}_x) &=& -(\cP_{\rm em,L} - \tilde{\cP}_{\rm em})
  \\ \noalign{\medskip}
    (S_{xR} - \tilde{S}_x) &=& +(\cP_{\rm em,R} - \tilde{\cP}_{\rm em})
  \end{array}
\end{equation}
where $\tilde{\cP}_{\rm em}$ is the electromagnetic pressure.

\subsection{Inner Solver}
%

As discussed before, the frozen limit allows us to employ any hydrodynamic relativistic solver inside the inner fan.
In order to capture the three-wave pattern characterizing the actual solution, our method of choice is based on the approximate HLLC solver of \cite{MB2005}.
Writing explicitly the jump conditions across $\lambda_L$ and $\lambda_R$ in the $x$-direction (similar results are obtained by index permutation) and taking advantage of the fact that EM fields do not experience any jump inside the inner fan, one finds
\begin{equation}\label{eq:HLLC_inner_jumps}
  \lambda_S(\tilde{\cW}_S - \cW^*_S) = \tilde{\cH}_S - \cH^*_S
\end{equation}
where $S=L,R$ and $\cW = (D,\, Q_x,\, Q_y,\, Q_z, \cE_h)$, $\cH = (Dv_x,\, Q_xv_x + p,\, Q_yv_x,\, Q_zv_x,\, Q_x)$ are the subset of  $\cU$ and $\cF$ containing only hydrodynamical quantities.
Notice that, by virtue of Eq. (\ref{eq:inner_mx+E}), EM quantities have disappeared from the jump conditions and that the tilde sign can be dropped since since hydrodynamical variables are continuous across the outer waves, that is, $\tilde{\cW}_{L,R} = \cW_{L,R}$ and $\tilde{\cH}_{S} = \cH_{S}$ for $S=L,R$.

The system of equations (\ref{eq:HLLC_inner_jumps}) is thus identical to Eq. [16] of \cite{MB2005} and can be solved likewise by imposing continuity of pressure and normal velocity across the contact wave, $p^* = p^*_L = p^*_R$ and $\lambda^* = v^*_{x,L} = v^*_{x,R}$.
To this purpose, we introduce the average state and flux
\begin{equation}
  \cW^{\rm hll} = \frac{\lambda_R\cW_R - \lambda_L\cW_L + \cH_L - \cH_R}
                       {\lambda_R - \lambda_L}
\end{equation}
and
\begin{equation}
  \cH^{\rm hll} = \frac{\lambda_R\cH_L - \lambda_L\cH_R
                         + \lambda_L\lambda_R(\cW_R - \cW_L)}
                         {\lambda_R - \lambda_L} \,,
\end{equation}
which allow us to rewrite the jump conditions (\ref{eq:HLLC_inner_jumps}) across $\lambda_L$ and $\lambda_R$ equivalently as
\begin{equation}\label{eq:HLLC_inner_jumps2}
    \lambda_S(\cW^{\rm hll} - \cW^*_{S}) = \cH^{\rm hll} - \cH^*_{S}
\end{equation}
where, again, $S=L,R$.
Imposing the momentum-energy relation $Q_x = (\cE_h + p)v_x$ and combining the energy and $x-$component of the momentum jump conditions (\ref{eq:HLLC_inner_jumps2}) leads to a quadratic equation for $\lambda^*$:
\begin{equation}\label{eq:HLLCquadratic}
    \cH_{[\cE]}^{\rm hll}\left(\lambda^*\right)^2
  - \left(\cW_{[\cE_h]}^{\rm hll} + \cH^{\rm hll}_{[Q_x]}\right)\lambda^*
  + \cW_{[Q_x]}^{\rm hll} = 0
\end{equation}
where $[.]$ denote a specific component of the array.
Eq. (\ref{eq:HLLCquadratic}) has one physical admissible solution given by the negative branch.
Once $\lambda^*$ is obtained, gas pressure in the star region is obtained from the momentum-energy equations as
\begin{equation}
  p^* = \cH^{\rm hll}_{[Q_x]} - \lambda^* \cH^{\rm hll}_{[\cE_h]}
\end{equation}
The hydrodyamic contribution to the interface flux is then given by
\begin{equation}
  \cH^* = \left\{\begin{array}{lcl}
    \cH_L  & {\rm if} & \lambda_L > 0            \\ \noalign{\medskip}
    \cH_L + \lambda_L(\cW^*_L - \cW_L)
                   & {\rm if} & \lambda_L < 0 < \lambda^*  \\ \noalign{\medskip}
    \cH_R + \lambda_R(\cW^*_R - \cW_R)
                   & {\rm if} & \lambda^* < 0 < \lambda_R  \\ \noalign{\medskip}
    \cH_R  & {\rm if} & \lambda_R < 0   \,.
  \end{array}\right.
\end{equation}

The total flux is finally  given by the sum of the hydro and EM contributions, the later being computed with the outer solver:
\begin{equation}
  \cF^*_\xf = \left(\begin{array}{l}
   \cH^*  \\ \noalign{\medskip}
    0_{\times 6} \end{array}\right)
    +
  \left(\begin{array}{l}
    0                         \\ \noalign{\medskip}
    \tilde{\tens{T}}_{in}     \\ \noalign{\medskip}
    \tilde{S}_n               \\ \noalign{\medskip}
    \epsilon_{ink}\tilde{E}_k \\ \noalign{\medskip}
   -\epsilon_{ink}\tilde{B}_k
    \end{array}\right)
\end{equation}
where $n$ labels the direction of propagation of the discontinuity ($x$ in our example).
Finally, the wave speed estimate for the fastest and slowest speeds $\lambda_L$ and $\lambda_R$ is done as in \cite{MB2005}.

\section{Numerical Benchmarks}
\label{sec:tests}
%
%
%

We now present a number of standard numerical benchmarks aiming at verifying and assessing the robustness, accuracy and computational performance of our resistive CT relativistic scheme.
Unless otherwise stated, test calculations will be carried out using the ideal equation of state (\ref{eq:eos}) with $\Gamma=4/3$, the MHLLC Riemann solver and the van Leer limiter (\ref{eq:VL_lim}).

Since the maximum characteristic velocity is equal to the speed of light, the time step is easily determined by the relation
\begin{equation}
  \Delta t = C_aN_{\rm dim}\left(  \frac{\alpha_1}{\Delta x}
                                 + \frac{\alpha_2}{\Delta y}
                                 + \frac{\alpha_3}{\Delta z}\right)^{-1} \,,
\end{equation}
where $C_a$ is the Courant number, $N_{\rm dim}$ is the number of spatial dimensions and  $\alpha_m = 1$ if $N_{\rm dim} \ge m$ and $\alpha_m=0$ otherwise.
We set the default Courant number to $C_a=0.4$ for two-dimensional problems and $C_a=0.25$ for three-dimensional ones.

\subsection{Telegraph Equation}
%

\begin{figure*}
  \centering
  \includegraphics[width=0.7\textwidth]{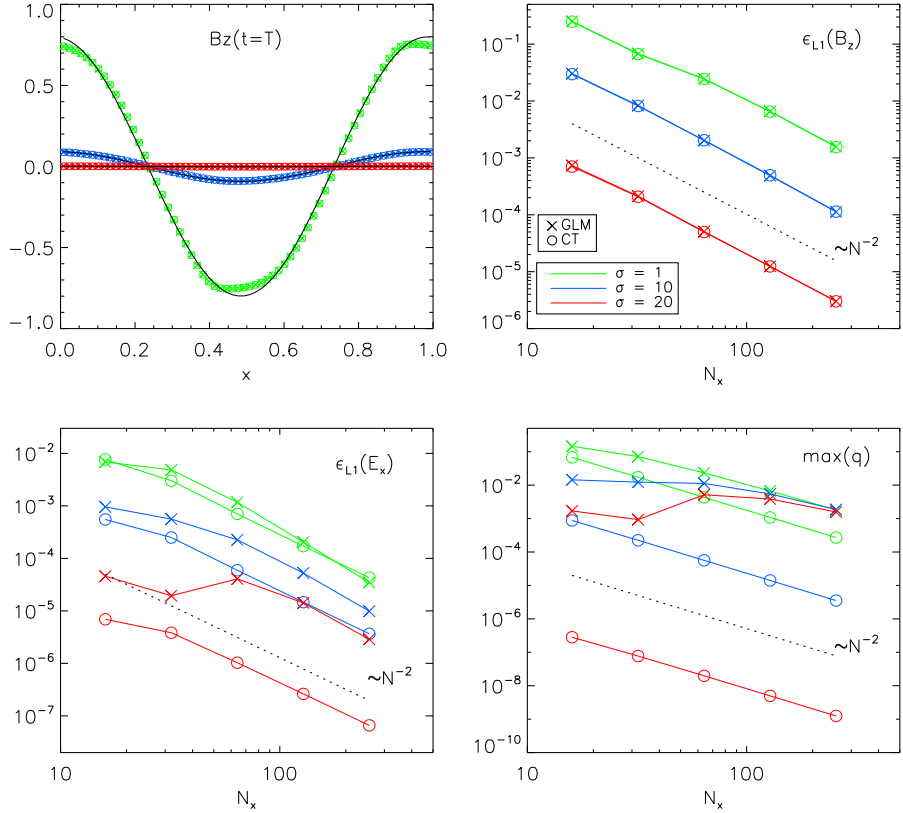}
  \caption{Telegraph equation test.
  \emph{Top left panel}: One-dimensional horizontal profiles (at $y=\Delta y/2$)
  of $B_z$ using different values of the conductivity $\sigma=1, 10, 20$
  corresponding to green, blue and red lines.
  The black line gives the exact solution after one period while results obtained
  with the CT and GLM schemes are shown using circles and plus signs.
  \emph{Top right panel}: L1 norm errors of the $B_z$ as functions of the
  resolution.
  \emph{Bottom panels}: L1 norm errors for the normal component of the electric
  field (left) and maximum value of the charge (right) as functions of the
  resolution.
  }
  \label{fig:telegraph_eq}
\end{figure*}

In the first test we consider the propagation of light waves in presence of finite value of conductivity.
Specifically, we want to solve the Maxwell's equation in the fluid rest frame,
\begin{equation}\label{eq:tg_Maxwell}
  \left\{\begin{array}{lcl}
    \DS \pd{\vec{B}}{t} + \nabla\times\vec{E} & = & 0 \\ \noalign{\medskip}
    \DS \pd{\vec{E}}{t} - \nabla\times\vec{B} & = & -\sigma\vec{E}  \,.
  \end{array}\right.
\end{equation}
Solutions of Eq. (\ref{eq:tg_Maxwell}) also satisfy the telegraph equation which is obtained upon differentiating the previous system with respect to time,
\begin{equation}\label{eq:telegraph}
  \left\{\begin{array}{lcl}
    \DS   \pd{^2\vec{B}}{t^2} + \sigma\pd{\vec{B}}{t} &=& \nabla^2\vec{B}
    \\ \noalign{\medskip}
    \DS   \pd{^2\vec{E}}{t^2} + \sigma\pd{\vec{E}}{t} &=& \nabla^2\vec{E}
         - \nabla q \,.
  \end{array}\right.
\end{equation}

The system of Maxwell's equation (\ref{eq:tg_Maxwell}) admits plane wave solutions with wavenumber $k$ and frequency $\omega$ tied by the dispersion relation
\begin{equation}\label{eq:tg_dr}
  \omega = -i\frac{\sigma}{2} \pm \mu \,,
  \qquad{\rm where}\qquad \mu = \sqrt{k^2 - \frac{\sigma^2}{4}} \,,
\end{equation}
which, as noted in \cite{MMB2018}, is also the dispersion relation for the telegraph equation.
Eq. (\ref{eq:tg_dr}) admits propagating modes when $\sigma < 2|k|$.
From the eigenvector of the system (\ref{eq:tg_Maxwell}), we obtain the exact solution for a single mode,
\begin{equation}\label{eq:tg_exact}
  \begin{array}{lcl}
    B_z &=&\DS B_1 e^{-\sigma t/2}\cos \phi(x,t)
    \\ \noalign{\medskip}
    E_y &=&\DS B_1 e^{-\sigma t/2}
               \left[  \frac{\mu}{k}     \cos\phi(x,t)
                      +\frac{\sigma}{2k} \sin\phi(x,t)\right]\,,
    \end{array}   
\end{equation}
where $\phi(x,t) = kx - \mu t$, while $B_1$ is the initial perturbation amplitude.
For $\sigma\ne 0$, Eq. (\ref{eq:tg_exact}) describes the propagation of monochromatic damped light waves with phase speed $v_\phi=\mu/k$ and thus with dispersive character.
Damping is suppressed in the limit $\sigma\to0$ (and thus $v_\phi\to1$) where we recover the classical non-dispersive light wave propagation.

For numerical purposes we consider here an inclined wave front by rotating the 1D solution around the $z$-axis by an angle $\alpha$.
The wavevector has orientation $\vec{k}= k_x(1,\tan\alpha,0)$, where $k_x=2\pi/L_x$ while $\tan\alpha = L_x/L_y$.
The computational domain is $L_x=1$, $L_y=1/2$ so that $\tan\alpha=2$.
Eq. (\ref{eq:tg_exact}) with $B_1=1$ and $\phi(x,0)=\vec{k}\cdot\vec{x}$ is used to initialize the electric and magnetic field inside the computational box.
Three dimensional vectors are then rotated according to 
\begin{equation}
  \{\vec{E},\vec{B}\}' = \left(\begin{array}{lll}
     \cos\alpha  & -\sin\alpha  &  0     \\ \noalign{\medskip}
     \sin\alpha  &  \cos\alpha  &  0     \\ \noalign{\medskip}
              0  &           0  &  1
  \end{array}\right)\{\vec{E},\vec{B}\}
\end{equation}
where a prime indicates quantities in the rotated frame.
Instead of solving just Maxwell's equation, we solve the full set of the resistive RMHD equations by prescribing a large fluid inertia ($\rho=10^{12}$) so that the fluid velocities become negligible and Maxwell's equations decouple from the remaining conservation laws.
Periodic boundary conditions are imposed everywhere.

Numerical results are shown in Fig. \ref{fig:telegraph_eq} for the CT (circles) and GLM (plus signs) schemes using different values of $\sigma = 1, 10, 20$, corresponding to green, blue and red coloured lines.
Computed profiles for $B_z$ are compared against the exact solution after one period $T=2\pi/\mu$ in the top left panel of Fig. \ref{fig:telegraph_eq} at the resolution of $64\times32$ grid zones.
Corresponding errors in L1 norm are plotted, as a function of the resolution $N_x=16,32, ..., 256$ ($N_y=N_x/2$), in the top right panel indicating that CT and GLM yields very similar results.

Although similar errors are produced in the $z$ component of magnetic field, results show significant differences by inspecting the normal component of the electric field.
Note that no charge should be produced during the evolution since $E_x$ - the normal component of the electric field in the unrotated frame - should vanish identically.
Nevertheless, propagation along an oblique direction (which is not the main diagonal) does not lead to perfect cancellation of multidimensional terms so that a non-solenoidal component of the current is generated at the truncation level of the scheme.
This best illustrated in the bottom panels of Fig. \ref{fig:telegraph_eq} where we plot the L1 norm errors of $E_x$ (bottom left) and the maximum value of the charge (bottom right).
Our CT method yields, overall, more uniform convergence with resolution when compared to GLM and the discrepancy between the two schemes worsen for larger values of $\sigma$.
While the charge remains below $10^{-6}$ with the CT scheme, the GLM appears to produce spurious divergence with poor convergence rates.
At the largest conductivity ($\sigma=20$, red lines), for instance, the two methods differ by over 3 orders of magnitude.

\subsection{Rotated Shock-Tube}
%

\begin{figure*}
  \centering
  \includegraphics[width=0.7\textwidth]{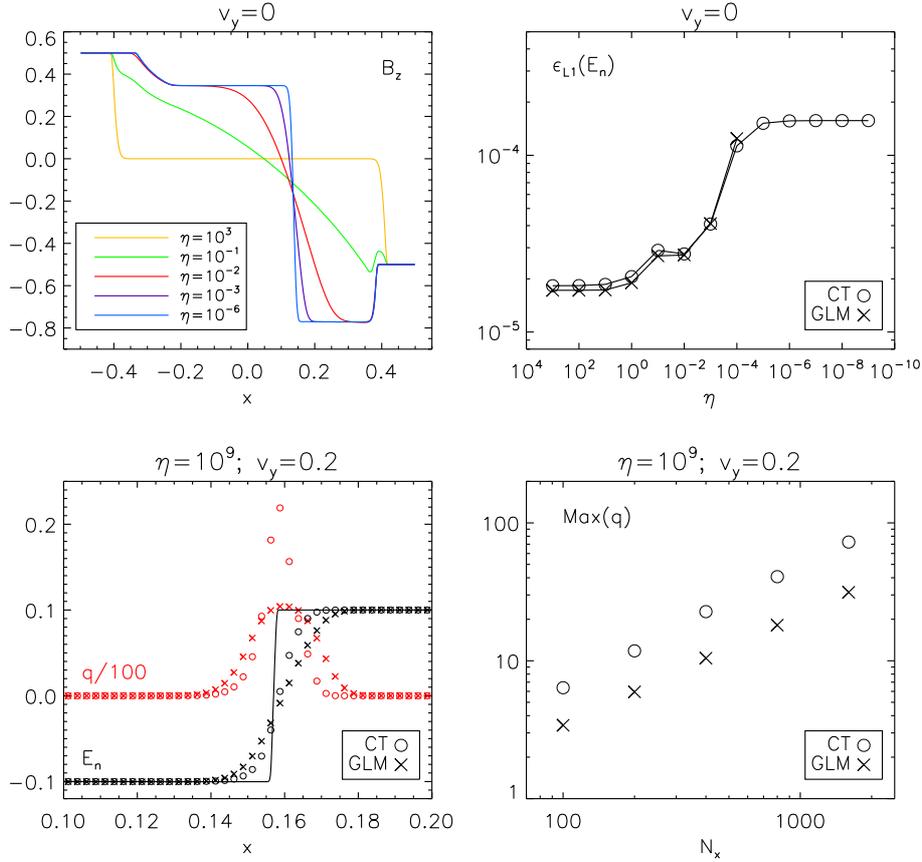}
  \caption{Results for the 2D rotated shock tube problem at $t=1/\sqrt{1+\tan^2\alpha}$.
           The $y$ component of the magnetic field (in the original unrotated
           frame) is plotted in the top-left panel for selected values of
           the $\eta$ as reported in the legend.
           In the top-right panel we plot the L1-norm errors of the normal component
           of electric field for the GLM and CT schemes as a function of $\eta$.
           In the bottom panels we show the normal component of $\vec{E}$ as well
           as the charge (red symbols, divided by $100$) for the second
           variant of the problem (left) and the maximum value of the charge as a function of the grid resolution (right).
           Results obtained with the CT and GLM schemes are shown using circles and
           crosses.
           }
  \label{fig:rotated_sod}
\end{figure*}
The shock-tube problem is a standard numerical benchmark consisting of an initial discontinuity separating two constant states.
Here we adopt a configuration similar to the one presented in \cite{Palenzuela_etal2009, Dionys_etal2013,Miranda_etal2018} and assign 1D left and right states as
\begin{equation}
  (\rho,\, p,\, B_z) = \left\{\begin{array}{ll}
  \DS \left( 1,\,
             1,\,
            \frac{1}{2} \right)      & \quad {\rm for}\quad x < 0\,,
  \\ \noalign{\medskip}
  \DS \left( \frac{1}{8},\,
            \frac{1}{10},\,
           -\frac{1}{2} \right)      & \quad {\rm for}\quad x > 0\,,
  \end{array}\right.
\end{equation}
while $B_x=B_y=0$ and the electric field is computed from the ideal condition, $\vec{E}=-\vec{v}\times\vec{B}$.
The ideal EoS (\ref{eq:eos}) with $\Gamma=2$ is used.

We consider two variants of the problem.
In the first one, the standard configuration with zero initial velocity is adopted: the electric field evolves by remaining perpendicular to the both the velocity and the magnetic field and the current density has a non-vanishing component only in the $y$-direction.
No charge is therefore produced during the evolution since $E_x=0$ holds at any time.
In the second variant, we prescribe an initial velocity $v_y=0.2$ everywhere so that the initial electric field presents a discontinuity at the origin and the charge is therefore $q(x) = -v_y(B_{zR} - B_{zL})\, \delta(x)$.

We rotate the initial condition by an angle $\alpha={\rm atan}(1/2)$ around the $z$ axis.
The computational domain is defined by a rectangular box of width $[-1/2,1/2]$ in the $x$-direction while in the transverse direction we have $y\in [-r/2, r/2]$ where $r=N_y/N_x$.
Zero-gradient boundary conditions hold at the rightmost and leftmost sides of the box, whereas for any flow variables $Q$ at the lower and upper boundaries we impose the translational invariance $Q(i,j)=Q(i\mp1, j\pm2)$.
Vectors are rotated accordingly as in the previous section.

In the first variant of this test, we set $N_x=400,\, N_y=8$ and stop computations at $t=0.4/\sqrt{1 + \tan^2\alpha}$ using different values of the resistivity in the range $\eta=10^3$ to $\eta=10^{-9}$ one decade apart.
The $z$-component of the magnetic field (unchanged under rotation) is plotted in the upper left panel of Fig. \ref{fig:rotated_sod} for the same values of $\eta$ also considered by \cite{Palenzuela_etal2009}.
In the top right panel of the same figure we plot the error (in L1 norm) of the normal component of the electric field $E_n = E_x\cos\alpha + E_y\sin\alpha$ obtained with our CT schemes as well as with  the GLM scheme for the entire range of values of $\eta$.
While the errors are similar for the two schemes, integration with the GLM scheme was not possible for values of $\eta$ below $10^{-5}$ due to numerical instabilities.
Errors increase sharply at around $\eta\approx 10^{-4}$ as the solution becomes progressively steeper and the scheme accuracy asymptotically reduces by one order of magnitude.
For smaller values of $\eta$, numerical resistivity becomes comparable and no difference can be noticed at this resolution.

In the second variant of this test, the resistivity has been fixed to a large value ($\eta = 10^9$) and a jump in the transverse velocity component is also present.
A non-zero delta-spike charge appears since the normal component of $\vec{E}$ is now discontinuous.
Both $E_n$ and the charge are advected at the local fluid velocity as shown in the bottom left panel of Fig. \ref{fig:rotated_sod} where a closeup view in the range $x\in[0.1,0.2]$ is plotted. 
Our CT method yields a sharper representation of the discontinuity and the value of the charge is twice the value obtained with GLM (the exact value being infinitely large).
Last, in the bottom-right panel, we show the maximum value of the charge density for different grid resolutions $N_x=100,200,400,800,1600$.

\subsection{Magnetized Blast Wave}
%
%
%
%
%
%

\begin{figure*}
  \centering
  \includegraphics[width=0.9\textwidth]{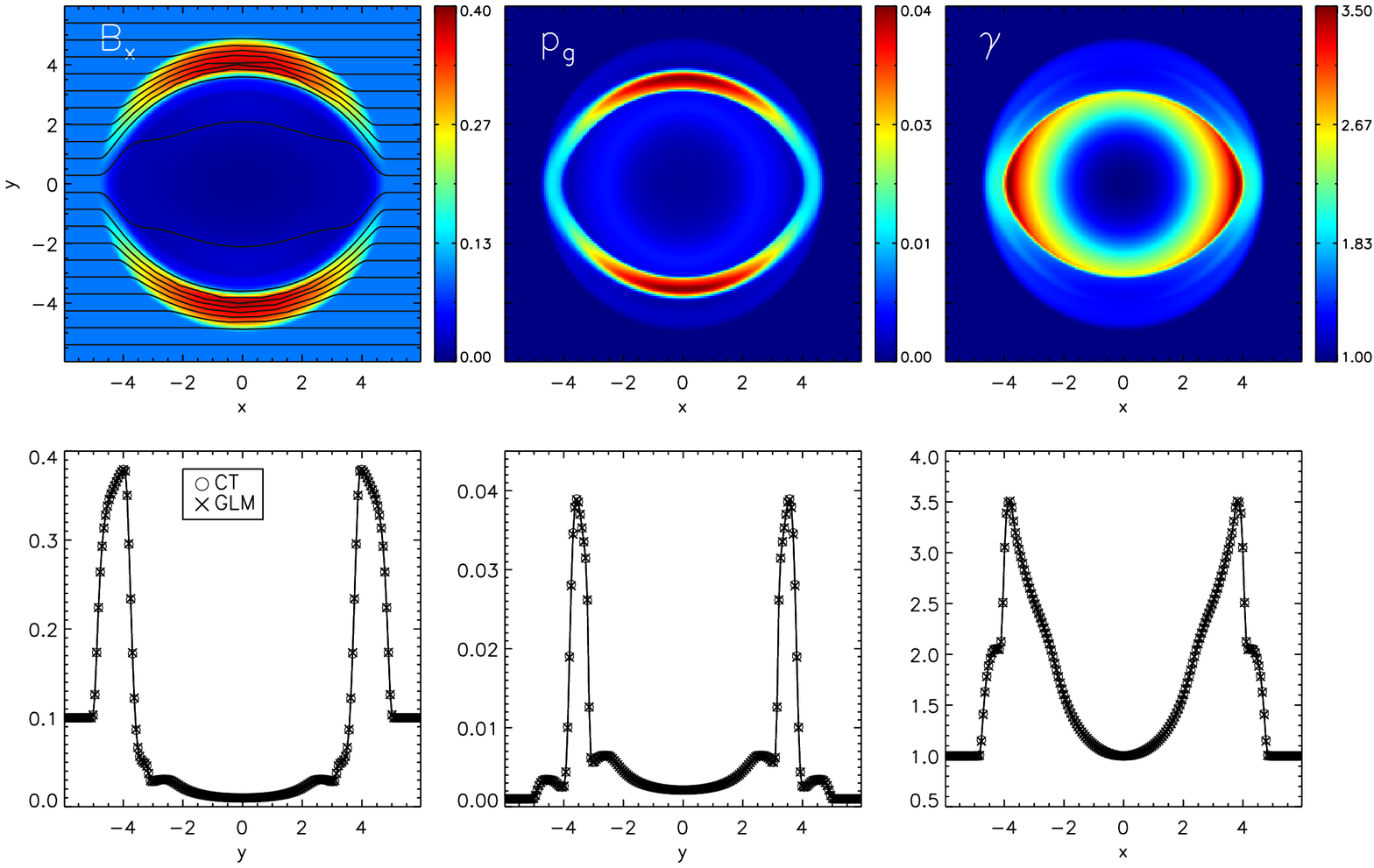}
  \caption{Cylindrical explosion at $t=4$ using $\eta=10^{-6}$. 
           Upper panels show, from left to right, the $x$ component of
           magnetic field, gas pressure and Lorentz factor computed using
           the CT scheme and the MHLLC Riemann solver.
           In the lower panels we plot the corresponding profiles along the
           $y$ axis (for $B_x$ and $p_g$) or the $x$ axis (for the Lorentz factor).
           Results obtained with the CT and GLM schemes are shown using circles
           and $x$ symbols.}
  \label{fig:blast2D}
\end{figure*}

The magnetized blast wave problem consists of a cylindrical (in 2D) or spherical (in 3D) explosion propagating in a uniform magnetized medium.
It is mainly used to test the ability of the scheme in handling oblique shock waves propagating in strongly magnetized environments as well as the fidelity in preserving the symmetric/antisymmetric properties of the solution.
Potential flaws in the numerical scheme may easily lead to unphysical densities or pressures if the divergence-free condition is not properly controlled and the scheme does not introduce adequate dissipation across oblique discontinuous features \citep[see, for instance,][and references therein]{MB2006,MigTze2010}.

\subsubsection{Cylindrical Blast Wave}
\label{sec:Blast2D}

Two-dimensional versions of this problem have been considered by previous investigators, e.g., \cite{Komissarov2007a, Palenzuela_etal2009, Mizuno2013, Dionys_etal2013, Miranda_etal2018}.
Computations are carried out inside the Cartesian square $x,y\in[-6,6]$ filled with uniform density and pressure $\rho = p = 10^{-3}$ for the exception of a small circular region, for $r\leq0.8$, where density and pressure take the values $\rho = 10^{-2}$ and $p = 1$.
Here $r=\sqrt{x^2+y^2}$ is the cylindrical radius.
For $0.8\leq r \leq 1$ continuity with the external environment is imposed through an exponential decay.
Velocity and electric field are initialized to zero, while the magnetic field is oriented along the $x$-direction $\vec{B} = (0.1,0,0)$ \citep[this is the configuration adopted by][]{Komissarov2007a, Miranda_etal2018}.
Computations have been performed using $N_x\times N_y = 200\times200$ grid zones and the system is evolved until $t = 4$.
Zero-gradient boundary conditions have been imposed on all sides.

Results obtained with the CT algorithm are shown in the top panels of Fig. \ref{fig:blast2D} where we display two-dimensional maps of the normal component of magnetic field, gas pressure and Lorentz factor.
The explosion produces a fast forward shock that propagates (nearly) radially leaving behind a reverse shock that delimits the inner region where expansion takes place radially.
Magnetic field lines pile up in the $y$-direction building up a shell of higher magnetic pressure.
The gas moves preferably in the $x$-direction where it achieves a higher Lorentz factor ($\gamma_{\max} \approx 3.62$).
Electric field and current have a non-vanishing component only in the $z$-direction and no charge is produced as $\nabla\cdot\vec{J}=0$ trivially holds.
We have checked our results with the GLM scheme and found no appreciable differences, as confirmed by the one-dimensional profiles along the $x$- and $y$-directions in the corresponding bottom panels.

The computational overhead brought by the CT scheme, which is intrinsically multidimensional, is partially balanced out by the lower number of variables to be evolved ($14$ vs $11$).
For this particular 2D configuration, for instance, we found that our CT method is approximately $\sim 5\%$ more expensive than the GLM scheme.

\subsubsection{Spherical Blast Wave}
\label{sec:Blast3D}

\begin{figure*}
  \centering
  \includegraphics[width=0.9\textwidth]{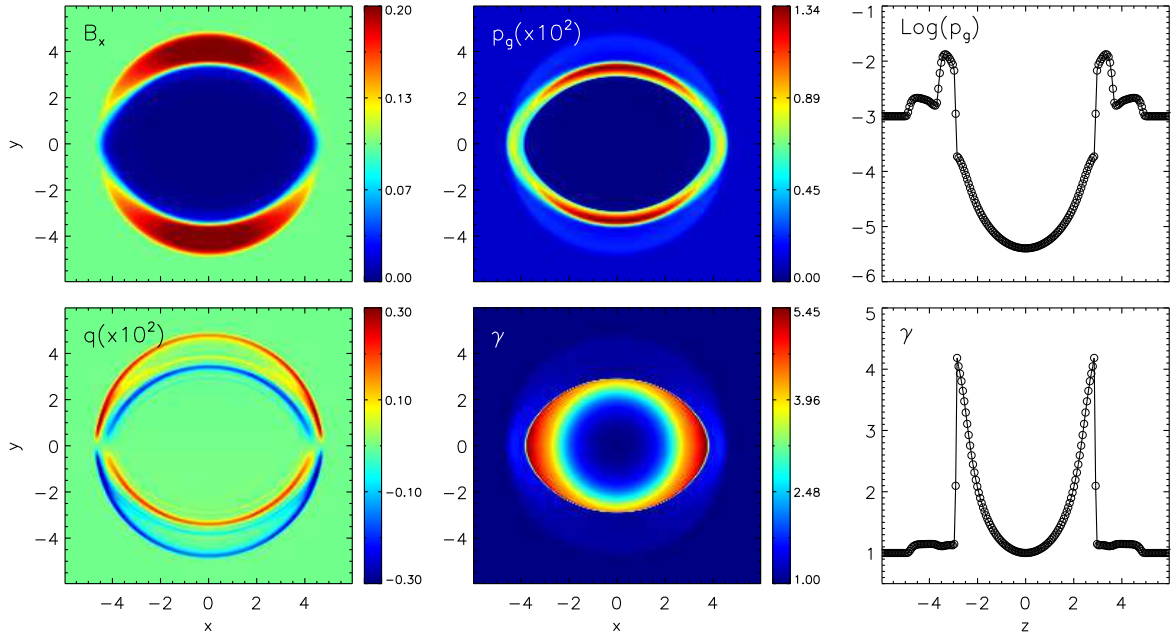}
  \caption{Spherical explosion at $t=4$ using $\eta = 10^{-6}$ and the CT scheme.
           The upper panels show, from left to right, 2D slices in the $xy$  plane of
           $B_x$, gas pressure $p_g$ and its 1D profile along the $z$ axis. 
           In the bottom panels we show 2D slices of the charge,
           Lorentz factor together with its one-dimensional cuts of along
           the $z$ axis.}
  \label{fig:blast3D}
\end{figure*}

The three-dimensional version of this problem has been formerly examined by \cite{Komissarov2007a, Dionys_etal2013} by extending the previous configuration to a square Cartesian box $x,y,z \in[-6,6]$ with initial conditions identical to the 2D case but with $r=\sqrt{x^2+y^2+z^2}$ now being the spherical radius.
Parameters are the same as for the cylindrical explosion.
In particular, note that our initial magnetic field is $\vec{B} = 0.1\hvec{e}_x$ and thus twice the value used by \cite{Dionys_etal2013}.
The same configuration is used in the original work by \cite{Komissarov2007a} although  we adopt here a much smaller value of resistivity, $\eta = 10^{-6}$.

Fig. \ref{fig:blast3D} shows 2D slices of the solution in the $xy$ plane at $t=4$ obtained with the CT scheme at the resolution of $200^3$ grid zones.
The solution is qualitatively similar to the 2D case although a few differences are worth noticing.
The inner region is delimited by a stronger reverse shock ($\gamma_{\max}\approx 5.5$)  and encloses a stronger rarefaction wave when compared to the 2D case where gas pressure reaches much smaller values, $\min(p) \approx 4\times10^{-6}$.
The plasma is accelerated mostly in the $x$-direction to larger Lorentz factor, $\gamma\approx 5.5$, when compared to the 2D case ($\gamma\approx3.5$).
Another crucial difference is the local production of non-zero charge which was absent from the 2D case: this reveals an important difference between the stability and robustness of the two methods.
Although both CT and GLM conserve charge up to machine accuracy ($\sim 10^{-15}$ is the total integrated charge for the two methods), our CT method runs without any problem whereas  GLM failed already for values of the resistivity smaller than $10^{-2}$ (unless the time-step is lowered) owing to large-amplitude oscillations in the charge.

The relative computational cost between CT and GLM is larger in 3D than in 2D and, in our implementation, it reaches approximately $15\%$ for this particular case.
This owes to the increased operation count which, in the 3D staggered method, accounts not only for the spatial dimensionality but also for the additional spatial reconstructions required by the multidimensional Riemann solver.

\subsection{Stationary Charged Vortex}
%

\begin{figure*}
  \centering
  \includegraphics[width=0.8\textwidth]{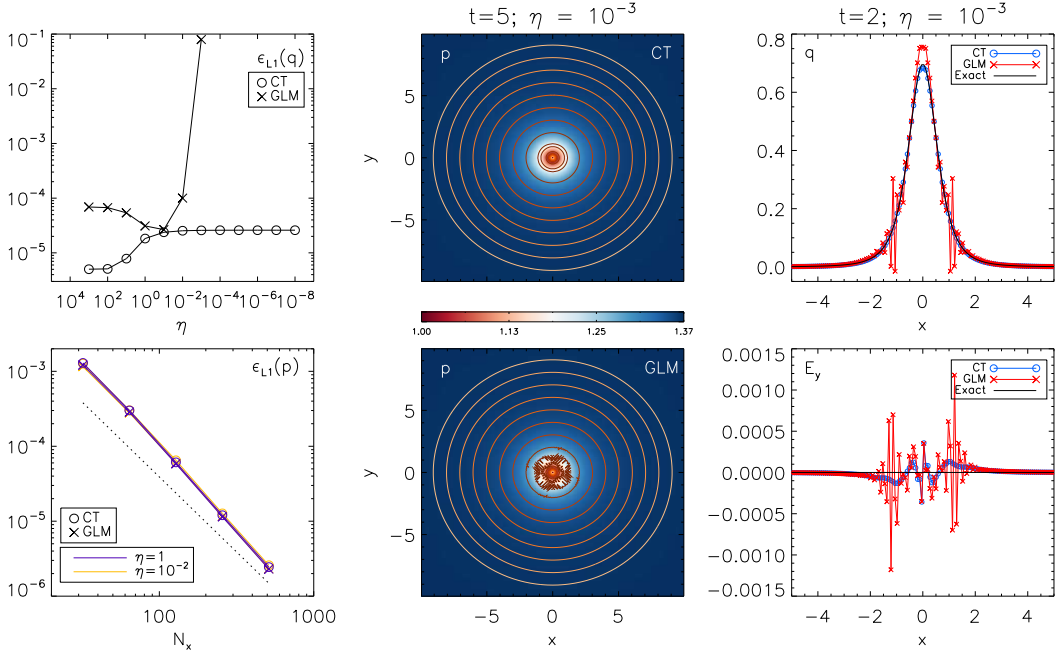}
  \caption{
   Numerical results of the charged vortex problem.
   \emph{Top left panel}: L1 norm errors of the charge as functions of the 
   resistivity using CT (circles) and GLM methods (X-symbols) with $256^2$ zones
   (integration with GLM was not possible for $\eta < 10^{-3}$).
   \emph{Bottom left panel}: L1 norm errors of the gas pressure as functions
   of the resolution. The dotted lines gives the expected $2^{\rm nd}$-order
   scaling.
   \emph{Central panels}: we show 10 equally-spaced contour levels of $|\vec{E}|$
   (chosen as $(q_0/2)r_k/(r_k^2 + 1.0)$, where $r_k = 1...9$) overlaid
   on the coloured maps of pressure at $t=5$.
   Resistivity is set to $\eta = 10^{-3}$, top and bottom panels corresponds
   to CT and GLM integrations with $256^2$ zones.
   \emph{Right panels}: 1D horizontal cuts at $y=0$ of charge and $E_y$ at
   $t=3$ obtained with CT (blue line with blue circles) and GLM
   (red line with X symbols).}
  \label{fig:charged_vortex}
\end{figure*}

We propose, for the first time to the extent of our knowledge, a new exact equilibrium solution of the fully Res-RMHD equations.
The solution is best described by adopting cylindrical coordinates $(r,\, \phi,\, z)$ and consists of a rotating flow with uniform density embedded in a vertical magnetic field $\vec{B}=(0,0,B_z)$.
In the ideal limit, this gives rise to a purely radial electric field, $\vec{E} = (E_r,0,0)$ with $E_r = -v_\phi B_z$.
Assuming purely radial dependence, the only non-trivial equations are the radial component of the momentum equation together with the $\phi$-component of Ampere's law:
\begin{align}
  \label{eq:charged_vortexA1}
  \DS  \left(\pd{p}{r} - \frac{w\gamma^2v_\phi^2}{r}\right) &=
  \DS  q E_r + qv_\phi B_z
    \\ \noalign{\medskip}
  \label{eq:charged_vortexA2}
  \DS  \pd{B_z}{r} & = -qv_\phi  \,,
\end{align}
where $q = \partial_r(rE_r)/r$ is the charge and we have used the fact $\vec{J} = (0, qv_\phi,0)$.
The appearance of a charge is a consequence of the fact that we now have a radial electric field generated by a rotating flow \citep[see, e.g., the review by][]{Spruit2013}.
Using the ideal electric field condition, $E_r+v_\phi B_z=0$, the right hand side of Eq. (\ref{eq:charged_vortexA1}) vanishes identically and the previous system of equations is rearranged as two independent ordinary differential equations, that is,
\begin{align}
  \label{eq:charged_vortexB1}
  \DS  \pd{p}{r} &= \DS \frac{w\gamma^2v_\phi^2}{r}
    \\ \noalign{\medskip}
  \label{eq:charged_vortexB2}
  \DS  \pd{H^2}{r} &= \frac{2}{r}E_r^2 \,,
\end{align}
where, using the same formalism already presented by \cite{Bodo_etal2013, Bodo_etal2016}, we have introduced $H^2(r) = B_z^2(r) - E_r^2(r)$.
The $H^2(r)$ function can be chosen arbitrarily provided the following conditions are met: i) $H^2(r) > 0$, which guarantees that velocity remain subluminal and ii) $dH^2(r)/dr(r=0)\ge 0$ which ensures that $E_r^2\ge 0$.
The equality sign holds at $r=0$ where the electric field must vanish.
A simple solution that satisfies the previous requisites is
\begin{equation}\label{eq:charged_vortexH2}
  H^2(r) \equiv B_z^2 - E_r^2 =  1 - \frac{q^2_0}{4} \frac{1}{r^2+1}
\end{equation}
where $q_0$ is the charge at $r=0$.
Using Eq. (\ref{eq:charged_vortexH2}), $E_r$ can be found by differentiating $H^2(r)$ with respect to $r$ (equation \ref{eq:charged_vortexB2}), $B_z$ and $v_\phi$ follow from Eq. (\ref{eq:charged_vortexH2}) and the ideal condition while gas pressure can be obtained by solving the differential equation (\ref{eq:charged_vortexB1}).
The final result is:
\begin{equation}\label{eq:charged_vortex}
  \left\{\begin{array}{l}
   E_r      = \DS \frac{q_0}{2}\frac{r}{r^2+1}
   \\ \noalign{\medskip}
   B_z      = \DS \frac{\sqrt{(r^2 + 1)^2 - q_0^2/4}}
                           {r^2 + 1} 
   \\ \noalign{\medskip}
   v_{\phi} = \DS -\frac{q_0}{2}\frac{r}{\sqrt{(r^2+1)^2 - q_0^2/4}}
   \\ \noalign{\medskip}
   p        = -\DS \frac{\rho}{\Gamma_1} + \left[
                   \frac{4r^2 + 4 - q_0^2}
                        {(r^2 + 1)(4 - q_0^2)}\right]^{\Gamma_1/2}
                   \left(p_0 + \DS\frac{\rho}{\Gamma_1}\right)
   \\ \noalign{\medskip}
   q        = \DS\frac{q_0}{(r^2+1)^2} \,,
  \end{array}\right.
\end{equation}
where $\Gamma_1 = \Gamma/(\Gamma-1)$.
We set density to unity while $p_0=0.1$ gives the pressure at infinity.
Charge is set to $q_0 = 0.7$.
Notice that the previous solution is also an exact solution of the ideal RMHD equations and that, using Eq. (\ref{eq:omegaB}), the rest-frame charge can be written as $q_0=-(\nabla\times\vec{v})\cdot\vec{B}$ at $r=0$.

We carry out computations on the two-dimensional Cartesian square $x,y\in[-10,10]$ using a uniform resolution of $N_x\times N_y$ zones and evolve the system until $t = 5$.
Boundary values are held fixed to the equilibrium solution throughout the integration.
Note that the equilibrium condition (\ref{eq:charged_vortex}) does not depend on the resistivity and numerical solutions carried out with different values of $\eta$ depend solely on the stability of the algorithm used for this particular problem.
This has been verified for a wide range of the resistivity parameter, namely, $\eta\in[10^3, 10^{-8}]$ using a grid resolution $N_x=N_y=256$.
Errors in L1 norm for the charge are plotted in the top left panel of Fig. \ref{fig:charged_vortex} as a function of $\eta$.
Our results confirm that the CT scheme remains stable for any value of the resistivity parameter in the chosen range.
In contrast, results obtained with GLM scheme give good agreement only for large values of $\eta$ while numerical instabilities is exhibited for $\eta \lesssim 10^{-3}$.
In the bottom left panel, we plot the errors (in $L1$ norm) of gas pressure as function of the resolution ($N_x = N_y = 32,64,128,256,512$) showing second-order convergence for both CT and GLM schemes.
Here $\eta = 1$ and $\eta = 10^{-2}$ have been used for the computations.

Numerical results for $\eta = 10^{-3}$, which at the resolution of $256$ zones sets the verge of stability for GLM, can be analyzed in the central panels of Fig. \ref{fig:charged_vortex} where we show coloured maps of pressure overlaid with iso-contour levels of the electric field magnitude (top and bottom).
Large oscillations are found in proximity of the coordinates origin with the GLM scheme (bottom central panel).
These large overshoots can also be recognized in the 1D horizontal cuts of charge $q$ and $E_y$ shown, respectively, in the top and bottom right panels using both CT (blue circles) and GLM (red X symbol) at $t=3$.
Note that the exact solution for the $y$-component of the electric field should vanish for $y=0$ but large-amplitude oscillations are clearly visible with GLM.
We point out that an increase in the resolution - which is accompanied by a reduction of the time step - extends the stability of the GLM method to lower values of the resistivity.
This result is consistent with the large errors introduced by the stiffness of the charge evolution equation.

\subsection{Tearing Mode}
%

\begin{figure*}
  \centering
  \includegraphics[width=0.9\textwidth]{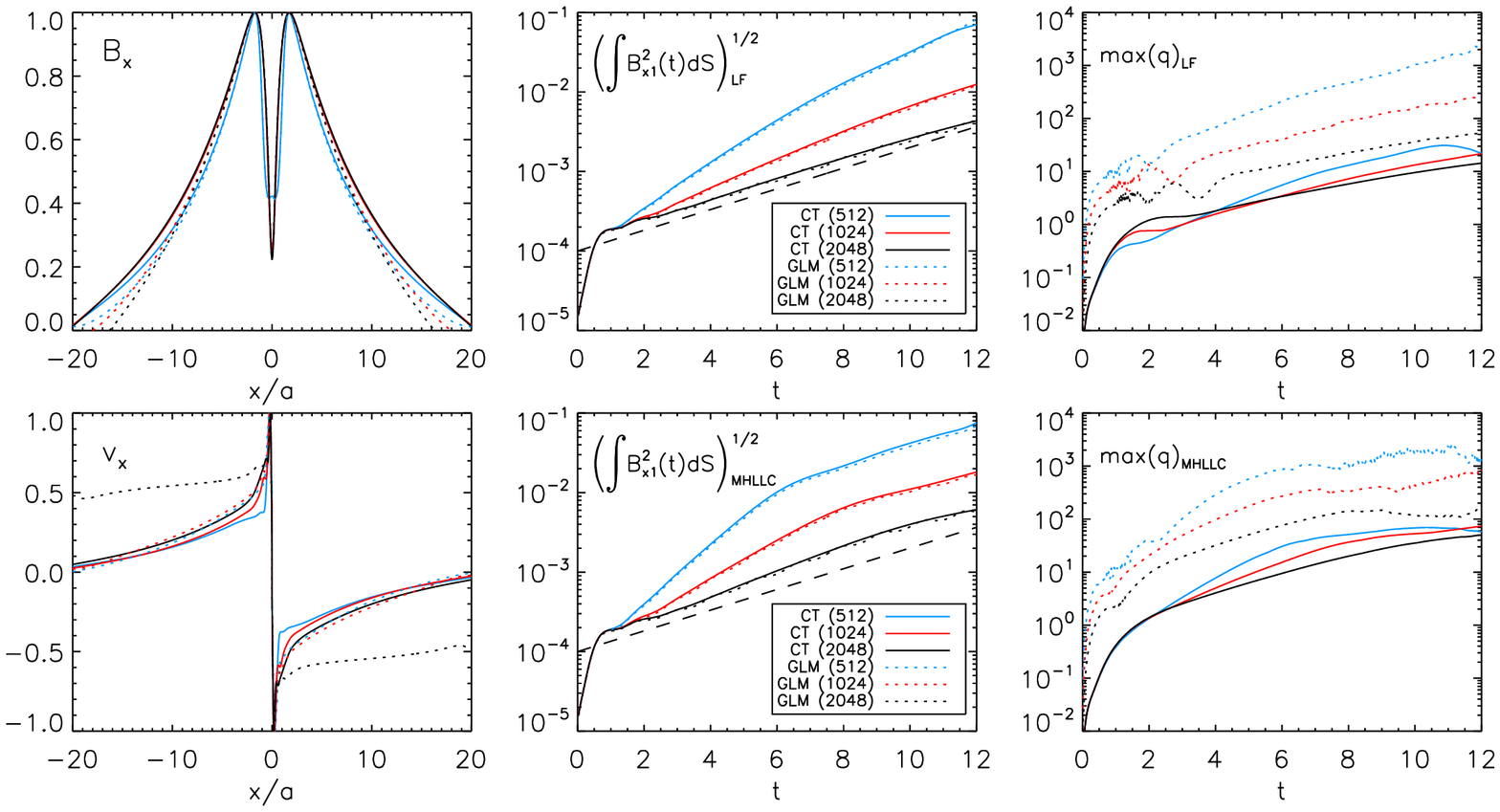}
  \caption{\emph{Left panels}: profiles of the $x$-components of magnetic field
           (top) and velocity (bottom) for the tearing mode instability at $t=10$
           using different methods and resolutions.
           The 1D cuts are taken along the $x$-axis for $y=0$ and $y=3/4L_y$, respectively
           and are normalized to unity.
           Solid (dotted) lines corresponds to the solutions obtained with
           the CT (GLM) methods while colors denote the different grid resolutions
           used $N_x\times N_x/4$ (green, red and black stand for
           $N_x=512, 1024, 2048$, respectively).
           Middle panels: measured growth rates for the selected cases using
           the LF solver (top) and the MHLLC solver (bottom).
           \emph{Right panels}: maximum value of the charge as a function of time.}
  \label{fig:tearing_mode}
\end{figure*}

\begin{figure*}
  \centering
  \includegraphics[width=0.9\textwidth]{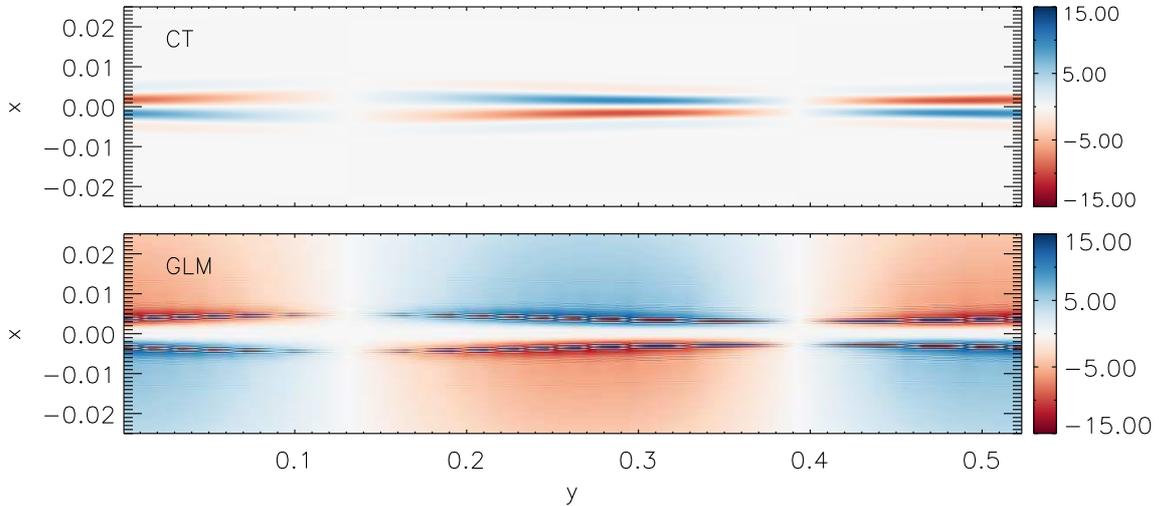}
  \caption{Colored distributions of the charge density obtained with the GLM (top)
           and CT (bottom) schemes at $t=10$ for $N_x=1024$.
           Both solutions have been generated with the LF Riemann solver.}
  \label{fig:tearing_mode_maps}
\end{figure*}

In the next example, we investigate the linear growth of the ideal tearing mode in a Harris-like current sheet, following the work of \cite{dZ_etal2016} and \cite{Miranda_etal2018}.
The initial configuration consists of an initially static ($\vec{v}=0$) and uniform plasma with constant density and pressure value, $\rho_0$ and $p_0$.
The initial magnetic field satisfies the force-free condition
\begin{equation}\label{eq:TM_background}
  \vec{B} = B_0\left[ \tanh\left(\frac{x}{a}\right)\hvec{e}_y
                     +\sech\left(\frac{x}{a}\right)\hvec{e}_z
               \right],
\end{equation}
where $a$ is the current sheet thickness. 
Useful parameters are the magnetization $\Sigma = B_0^2/\rho_0$, the plasma beta $\beta = 2p_0/B_0^2$, the Alfv\'en velocity $v_a = B_0/\sqrt{B_0^2 + w_0} = B_0/\sqrt{B_0^2 + \rho_0 + 4p_0}$ (for an ideal gas law with adiabatic index $\Gamma=4/3$), and the Lundquist number $S=v_a L /\eta \gg 1$, where $L$ is a typical spatial scale and $\eta$ the resistivity.
The electric field is initially zero everywhere.
According to the MHD works by \cite{Pucci_Velli2014, Landi_etal2015}, when extremely thin current sheets are considered the tearing mode ceases to be a slow process (with growth rate $\gamma_\mathrm{TM}\propto S^{-1/2}$) and reconnection occurs instead on the ideal Alfv\'enic time $\tau_a = L/v_a$. The threshold is provided by the critical (inverse) aspect ratio
\begin{equation}\label{eq:TM_ideal}
  a = S^{-1/3}L,  
\end{equation}
for which, in the asymptotic limit of large $S$ but independently on the actual value of $S$, $\gamma_{\rm TM}\simeq 0.6 v_a/L$. In the relativistic regime, where $v_a\to c=1$, this so-called \emph{ideal} tearing instability thus becomes a very efficient process. 

In order to trigger the instability we perturb the initial configuration with a single-mode magnetic field equal to
\begin{equation}\label{eq:tearing_perturb}
  \delta\vec{B} =  \epsilon B_0 \sech\left(\frac{x}{a}\right)
  \left(\begin{array}{l}
     \DS \cos(ky)                                            \\ \noalign{\medskip}
     \DS \frac{\sin(ky)}{ka}\tanh\left(\frac{x}{a}\right)    \\ \noalign{\medskip}
     0
  \end{array}\right)
\end{equation}
where $\epsilon = 10^{-4}$ is the initial perturbation amplitude and $k$ the wave number.
For computational purposes, we initialize the magnetic field in the $x-y$ plane using the $z$-component of the vector potential
\begin{equation}
A_z = -B_0 \left[     a \log\left(\cosh\frac{x}{a}\right)
                   - \frac{\epsilon}{k}\sin(ky)\sech\left(\frac{x}{a}\right)\right],
\end{equation}
which ensures an initial divergence-free discretization of $\vec{B}$ at the beginning.
Following \cite{dZ_etal2016}, we use for this test $L = B_0 = \Sigma = \beta = 1$, hence $\rho_0=1$, $p_0=0.5$, $v_a=0.5$ and $S=10^6$, so that $a=0.01$ and $\eta=5\times 10^{-7}$.
The theory predicts, for the \emph{ideal} tearing mode, a wave number for the fastest growing mode of $k_\mathrm{max}=1.4\, S^{1/6}=14$, providing $\gamma_\mathrm{TM}\simeq 0.3$.
However, probably due to the diffusion of the initial force-free field or to compressible effects, \cite{dZ_etal2016} found a maximum in the dispersion relation curve for $k=12$ and $\gamma_\mathrm{TM}\simeq 0.27$, hence, also following \cite{Miranda_etal2018}, we will adopt $k=12$ as the standard test value for the wave vector.

We perform computations on the rectangular domain $x\in[-20a,+20a]=[-0.2,+0.2]$ and $y\in[0,2\pi/k]=[0,0.5236]$, using free outflow conditions at the $x$ boundaries and periodicity along $y$.
For the sake of comparison, we have repeated calculations using our CT implementation as well as the GLM scheme with different grid resolutions corresponding to $N_x\times N_x/4$ zones with $N_x=512, 1024, 2048$.
The monotonized central limiter was used for these computations.
We note that GLM requires, for stability, a smaller CFL number $C_a=0.2$, whereas computations with CT are carried at twice this value, $C_a=0.4$.

In the left panels of Fig. \ref{fig:tearing_mode} we show horizontal cuts of the $x$-components of magnetic field and velocity at $t=10$ using CT (solid lines) and GLM (dashed lines) at the chosen grid size using different colors.
At the resolutions of $Nx=1024, 2048$ the profiles well approximate the eigenfunctions given by \cite{dZ_etal2016} (see their Figure 1).
Overall, the CT scheme performs similarly to the GLM method albeit with slightly reduced numerical dissipation and better convergence with resolution.

\begin{table}
 \centering
 \caption{Growth rates for the tearing mode instability measured from the simulations.
          Twelve cases are shown corresponding to three different grid resolutions
          (left column) and to different combination of divergence control schemes
          (GLM and CT) and Riemann solvers (LF and MHLLC).}
 \label{tab:tearing_mode}
  \begin{tabular}{|c|c|c|c|c|c|}
    \hline
                  &   \multicolumn{2}{ c }{GLM} &  & \multicolumn{2}{ c }{CT} \\
                  \cline{2-3}    \cline{5-6}  \\ 
    Resolution        &  LF    &  MHLLC  &   &   LF &  MHLLC \\
    \hline
$ 512 \times 128$  &    0.49  &    0.31 &   &    0.47  &    0.32 \\ 
$ 1024 \times 256$  &    0.36  &    0.31 &   &    0.36  &    0.31 \\ 
$ 2048 \times 512$  &    0.28  &    0.30 &   &    0.28  &    0.30 \\
\hline
  \end{tabular}
\end{table}
Perturbations are expected to grow exponentially as $Q_1\propto e^{\gamma_{TM}t}$ and we have measured the numerical growth rate by considering, as suggested by \cite{Miranda_etal2018}, the integral of the $x$-component of magnetic field (squared),
\begin{equation}\label{eq:TM_ft}
  f(t) = \frac{1}{2} \log\left(\int B_{x}^2(t) \, dS\right) \,.
\end{equation}
Plots of $f(t)$ are shown in the central panels of Fig. (\ref{fig:tearing_mode}) for the LF (top) and the MHLLC (bottom) Riemann solvers.
Our results indicate that increasing the resolution leads to smaller growth rates, in agreement with the previous findings.
However, two distinct phases can be discerned using the MHLLC solver: a steeper growth for $t \lesssim 6-8$ followed by a softer one for $t \gtrsim 6-8$, the actual value depending on the resolution.
The behavior remains unaltered when switching from CT to GLM and it is not observed by \cite{dZ_etal2016} and \cite{Miranda_etal2018} who used $5$-th or higher-order reconstructions.
For our second-order scheme, instead, we attribute this behavior to compressible effects enhanced by the resolution of density jumps, probably triggering spurious modes that grow faster in the early stage of evolution.
For this reason, the growth rates are computed as the slope obtained by the linear fit of Eq. (\ref{eq:TM_ft}) versus time over the interval $6\le t \le 12$ and reported in Table \ref{tab:tearing_mode}.
Results with the MHLLC solver show better convergence rates with resolution when compared to the more diffusive LF scheme (only few percent with resolution doubling).
At the largest resolution, the MHLLC Riemann solver yields larger growth rates than LF and converges to the actual value ($\gamma_{\rm TM}\approx 0.30$).

It is also instructive to compare the charge evolution obtained with the GLM and CT methods, as shown in the right panels of Fig. \ref{fig:tearing_mode} where we plot the maximum value of the charge versus time.
With the GLM scheme, a systematic excess of charge is produced which is noticeably reduced   by doubling the resolution (approximately one order of magnitude with the LF solver).
Fluctuations with the CT scheme are much less affected by the grid size and are restrained within a factor of 2.
Finally we show, in Fig. \ref{fig:tearing_mode_maps}, the charge density distribution obtained with the GLM and CT schemes at $t=10$ by narrowing the view down to the $x$-axis.
With the CT method, charge density is mostly concentrated around the current sheet and the solution appears to be well-behaved and oscillation-free.
On the contrary, with the GLM scheme, charge distribution spreads out on the sides and the solution develops large overshoots as well as spurious oscillations which appear to have numerical origin.

\subsection{Kelvin-Helmholtz Flow}
%

\begin{figure*}
  \centering
  \includegraphics[width=0.85\textwidth]{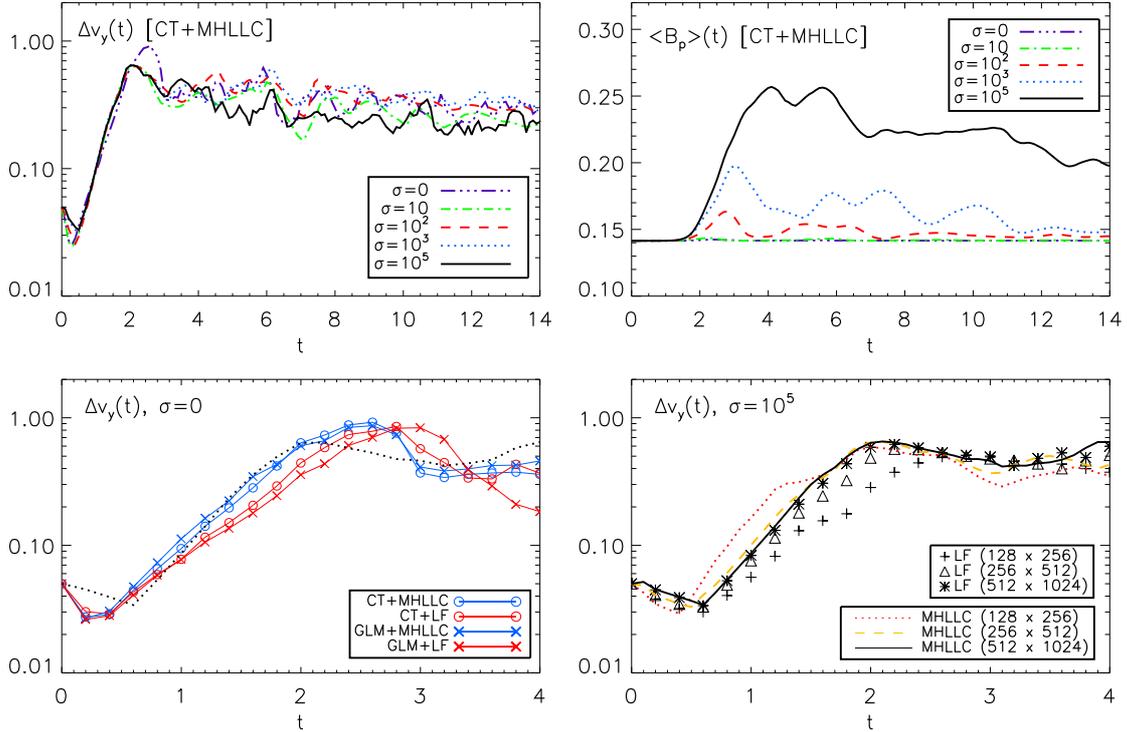}
  \caption{Growth rate and poloidal field amplification for the 
           resistive Kelvin-Helmholtz test problem.
           \emph{Top left panel}: measured growth rate as a function of time using
           selected values of conductivity (reported in the legend) at the
           resolution of $256\times512$ grid zones.
           \emph{Top right panel}: poloidal field amplification as a function
           of time.
           \emph{Bottom left panel}: growth rate at $\sigma=0$ comparing
           different schemes (CT and GLM with circles and cross symbols)
           and Riemann solvers (MHLLC and LF with red and blue lines,
           respectively).
           The dotted line gives the growth rate at $512\times1024$ zones
           and $\sigma=10^5$.
           \emph{Bottom right panel}: growth rate as function of the resolution
           comparing MHLLC with LF solvers (CT only).}
  \label{fig:kh_growth}
\end{figure*}

\begin{figure*}
  \centering
  \includegraphics[width=0.75\textwidth]{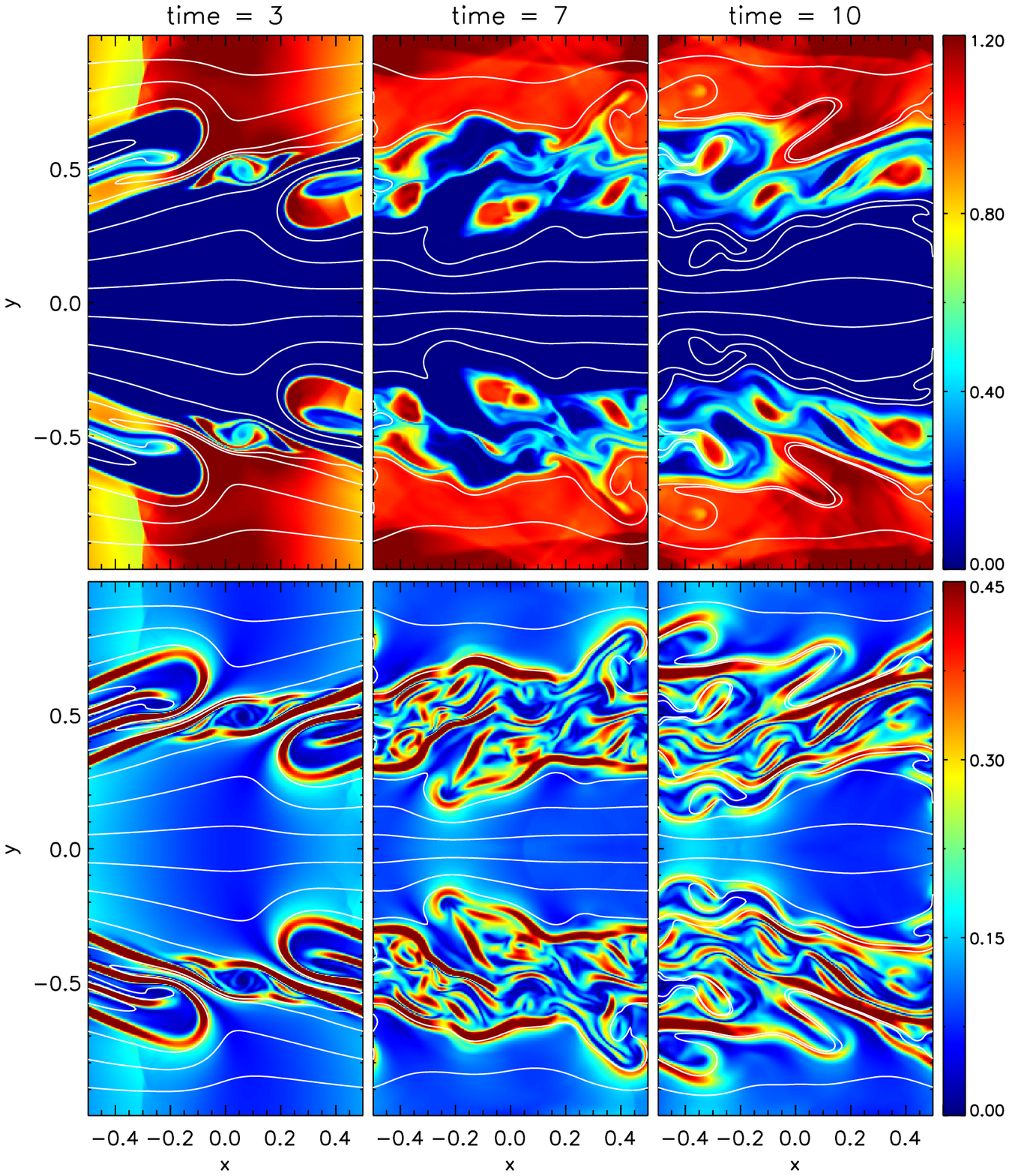}
  \caption{Time evolution of the Kelvin-Helmholtz problem for $\sigma=10^5$
   using the CT scheme and the MHLLC Riemann solver.
   Coloured distribution maps of density (top panels) and poloidal 
   to toroidal magnetic field ratio at time $t=3, 7$ and $10$.
   Magnetic field lines are overlaid.}
  \label{fig:kh_maps}
\end{figure*}

As a test application, we consider the evolution of a double shear layer as already presented, in the context of resistive relativistic flows by \cite{Mizuno2013} and in the ideal case by \cite{MUB2009, BeckStone2011}. 
The initial condition consists of a background (double) shear layer with non-uniform density distribution,
\begin{equation}
  \begin{array}{l}
  v_x  = \DS v_{\rm sh}\tanh\phi(y)
  \\ \noalign{\medskip}
  \rho = \DS   \frac{1+\tanh\phi(y)}{2}\rho_h
              + \frac{1-\tanh\phi(y)}{2}\rho_l  \,,
  \end{array}
\end{equation}
where $\phi(y) = (|y|-1/2)/a$.
The background flow is then perturbed by
\begin{equation}\label{eq:kh:vy}
  v_y  = \DS {\rm sign}(y) A_0v_{\rm sh}\sin(2\pi x)e^{-\phi(y)^2a^2/\alpha^2}\,.
\end{equation}
Following \cite{Mizuno2013} we take the velocity of the shear layer to be $v_{\rm sh} = 1/2$ with thickness $a=10^{-2}$ and the density contrast $\rho_h=1$, $\rho_l=10^{-2}$.
The parameters in the last equations, $\alpha = 10^{-1}$ and $A_0 = 0.1$, characterize the perturbation cutoff height and amplitude.
The magnetic field is initially constant and uniform with a poloidal ($B_x$) and toroidal ($B_z$, out of the plane) components parameterized by
\begin{equation}
  \vec{B} = \left(\sqrt{2p\mu_p},\, 0,\, \sqrt{2p\mu_t}\right) \,,
\end{equation}
where $\mu_p = 0.01$ and $\mu_t = 1$ are magnetization parameters.

The Cartesian box used for the integration of the resistive RMHD equations is defined by $x\in[-1/2,1/2]$, $y\in [-1,1]$ using different values of the conductivity, $\sigma = 1/\eta= 0, 10, 10^2, 10^3$ and $10^5$.
We perform computations using the MHLLC solver and the MC limiter (\ref{eq:MC_lim}) until $t=15$ with the nominal resolution of $256\times512$ grid zones.
Computations with the CT scheme remained stable at the nominal Courant number ($C_a=0.4$) for any value of the conductivity while numerical instabilities occurred at large $\sigma$'s with the GLM scheme unless the CFL number was lowered to $0.1$.

The growth rates, computed as $\Delta v_y = (\max(v_y) - \min(v_y))/2$, are shown in the top left panel of Fig. \ref{fig:kh_growth} for selected values of the conductivity.
Our measured growth rates favourably compare to those of \cite{Mizuno2013}, indicating that different values of the conductivity have a negligible impact on the growth of the instability.
The $\sigma = 0$ case (purple dashed-triple dotted line), which in the work by \cite{Mizuno2013} yielded a smaller growth rate, did not make particular difference in our case.
In all likelihood, this discrepancy can be attributed to the choice of the Riemann solver as it can be inferred from the bottom left panel of Fig. \ref{fig:kh_growth}, where GLM and CT schemes are compared using the LF (red) and MHLLC (blue) solvers.
Switching from the former to the latter leads to a larger growth rate which become closer to the high-conductivity case (black dotted line).
In the top right panel, we plot the poloidal field amplification which is enhanced with increasing conductivity in accordance with \cite{Mizuno2013}.
However, our five-wave solver leads to a steeper poloidal field amplification when compared to the scheme of \cite{Mizuno2013} and to an earlier start of the saturation phase \citep[$t\approx 4$ instead of $t\approx 5$, see Figure 11 of][]{Mizuno2013}. 
A resolution study, shown in the bottom right panel of Fig. \ref{fig:kh_growth} for $\sigma=10^5$, confirms that the MHLLC solver yields larger growth rates and faster convergence when compared to the simpler LF scheme \cite[similar results were found with the HLLD Riemann solver in the Kelvin-Helmholtz test presented by][]{MUB2009}.

The time evolution is shown in Fig. \ref{fig:kh_maps} for the CT scheme with $\sigma = 10^5$ at three different snapshots, $t=3,7$ (left and middle panels) and $t=10$ (right panel).
Top and bottom panels show, respectively, coloured distributions of density and the poloidal to toroidal magnetic field ratio ($\sqrt{B_x^2 + B_y^2}/B_z$).
The linear phase is followed by vortex formation and the transition to the nonlinear regime during which the mixing layer enlarges and magnetic field becomes amplified and stretched into filamentary structures.
Note also the formation of the intermediate vortex which does not appear when using the LF Riemann solver.

Computations with the CT scheme required approximately $5\%$ more time than the GLM case.
The same result was established for the 2D Blast wave test problem, see section \ref{sec:Blast2D}.

\subsubsection{Extension to Three-Dimensions}
%

\begin{figure}
  \centering
  \includegraphics[width=0.45\textwidth]{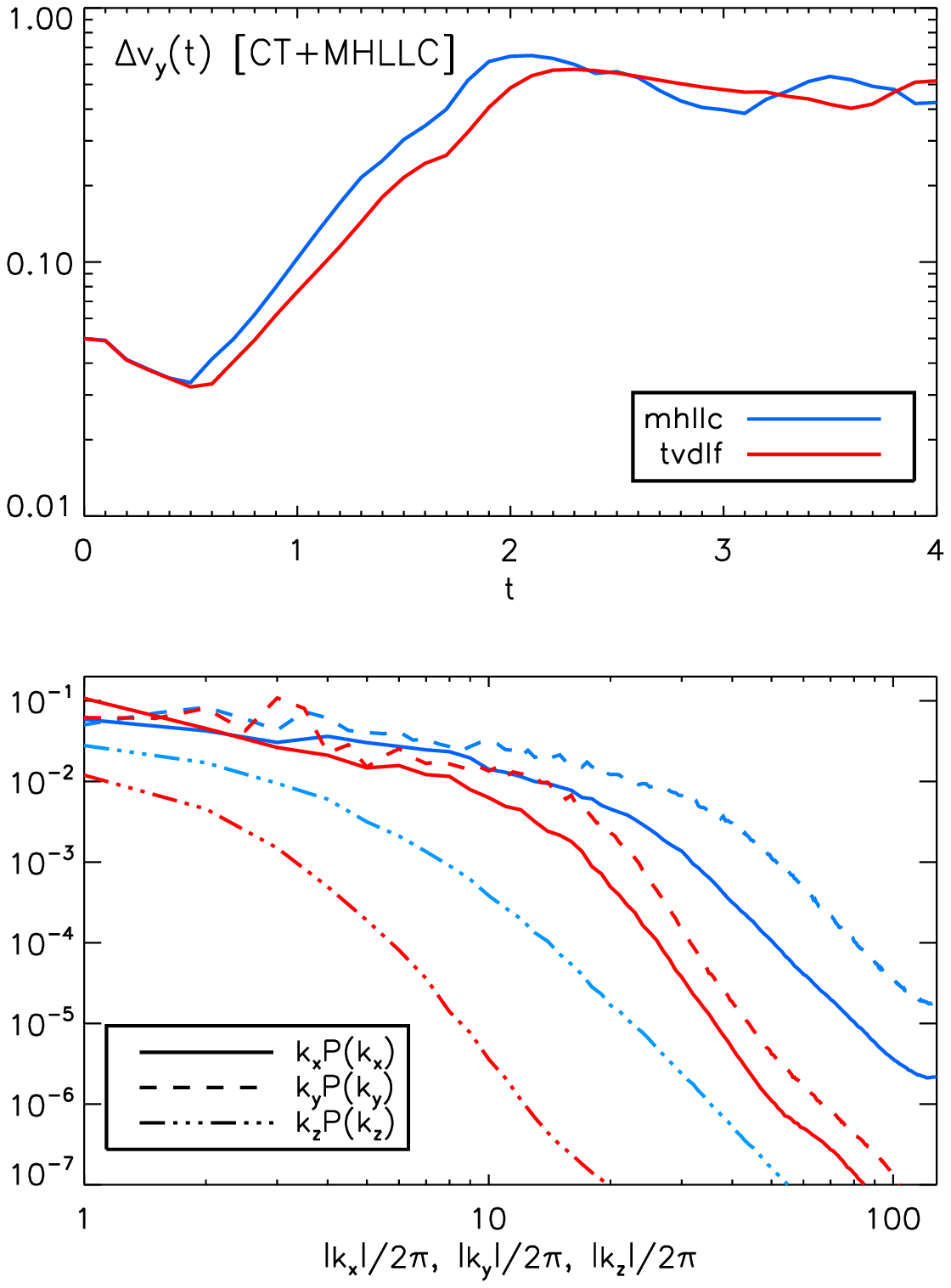}
  \caption{Growth rate as a function of time (top) and power spectra
           at $t=15$ (bottom) for the 3D Kelvin-Helmholtz instability.
           Blue and red lines correspond to the MHLLC and LF cases.
           In the bottom panel, integrated power spectra in the $x$, $y$
           and $z$ directions are plotted using solid, dashed and dashed-dotted
          line style.}
  \label{fig:kh3D_growth}
\end{figure}

\begin{figure*}
  \centering
  \includegraphics[width=0.3\textwidth]{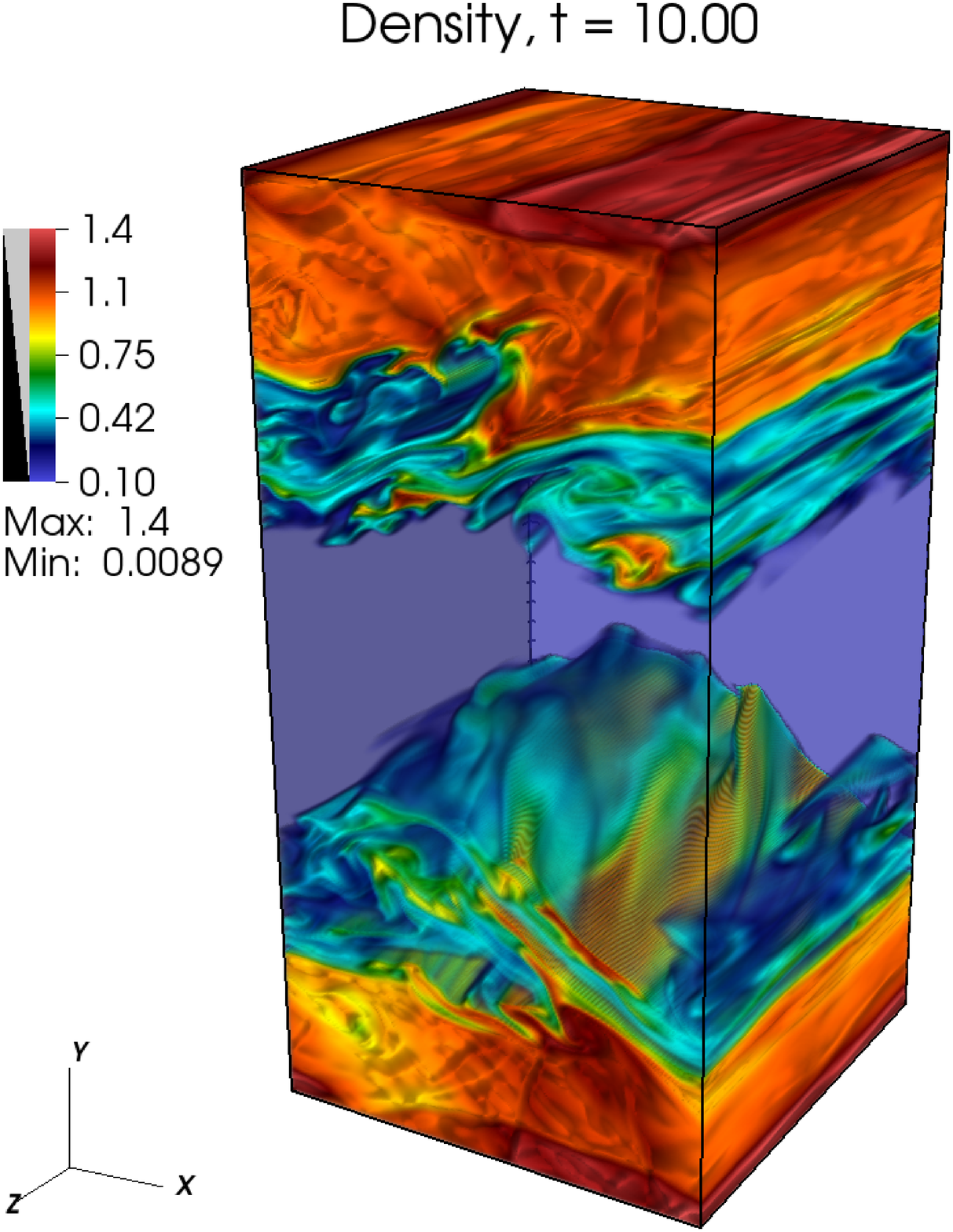}  \hspace{10pt}
  \includegraphics[width=0.3\textwidth]{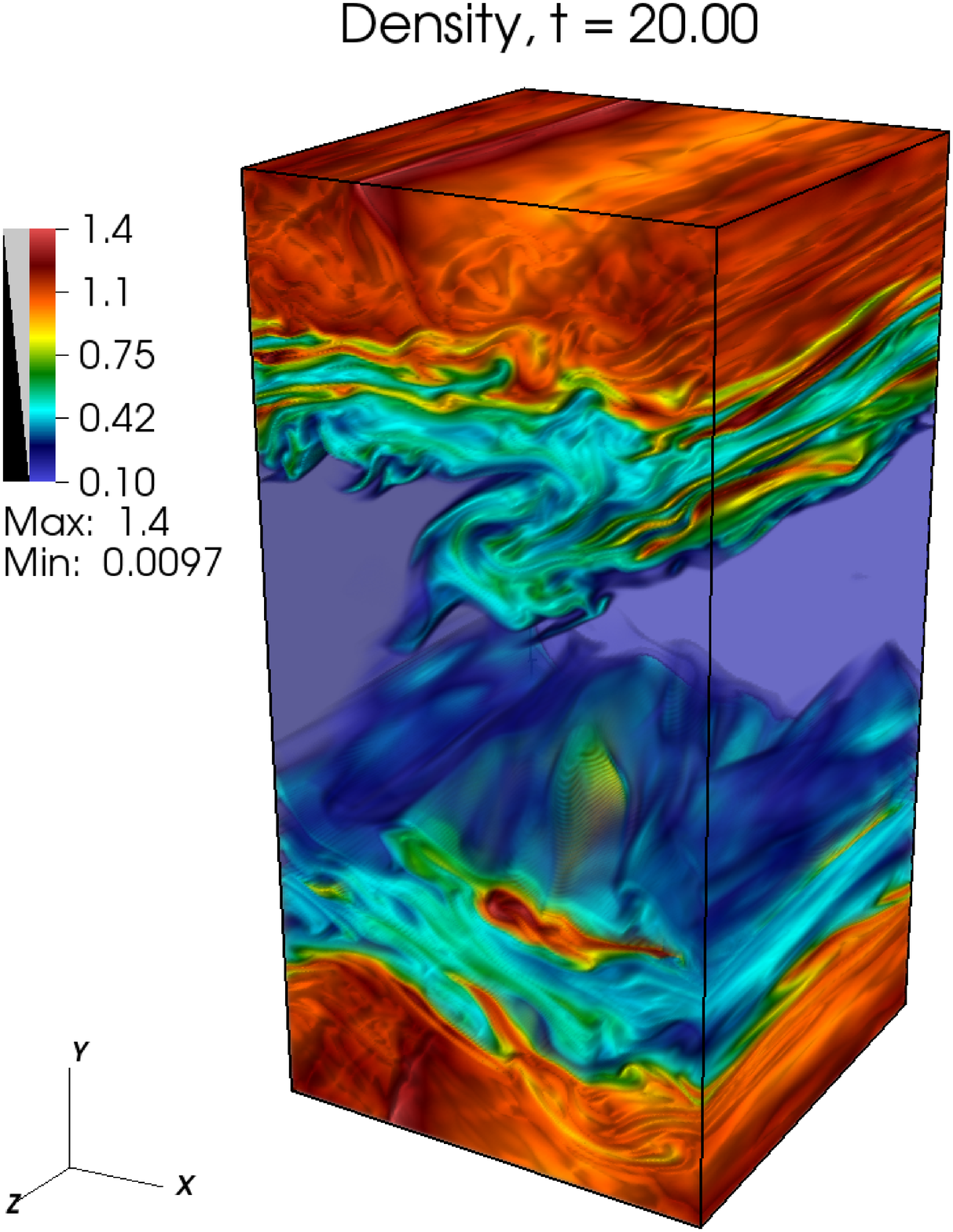}  \hspace{10pt}
  \includegraphics[width=0.3\textwidth]{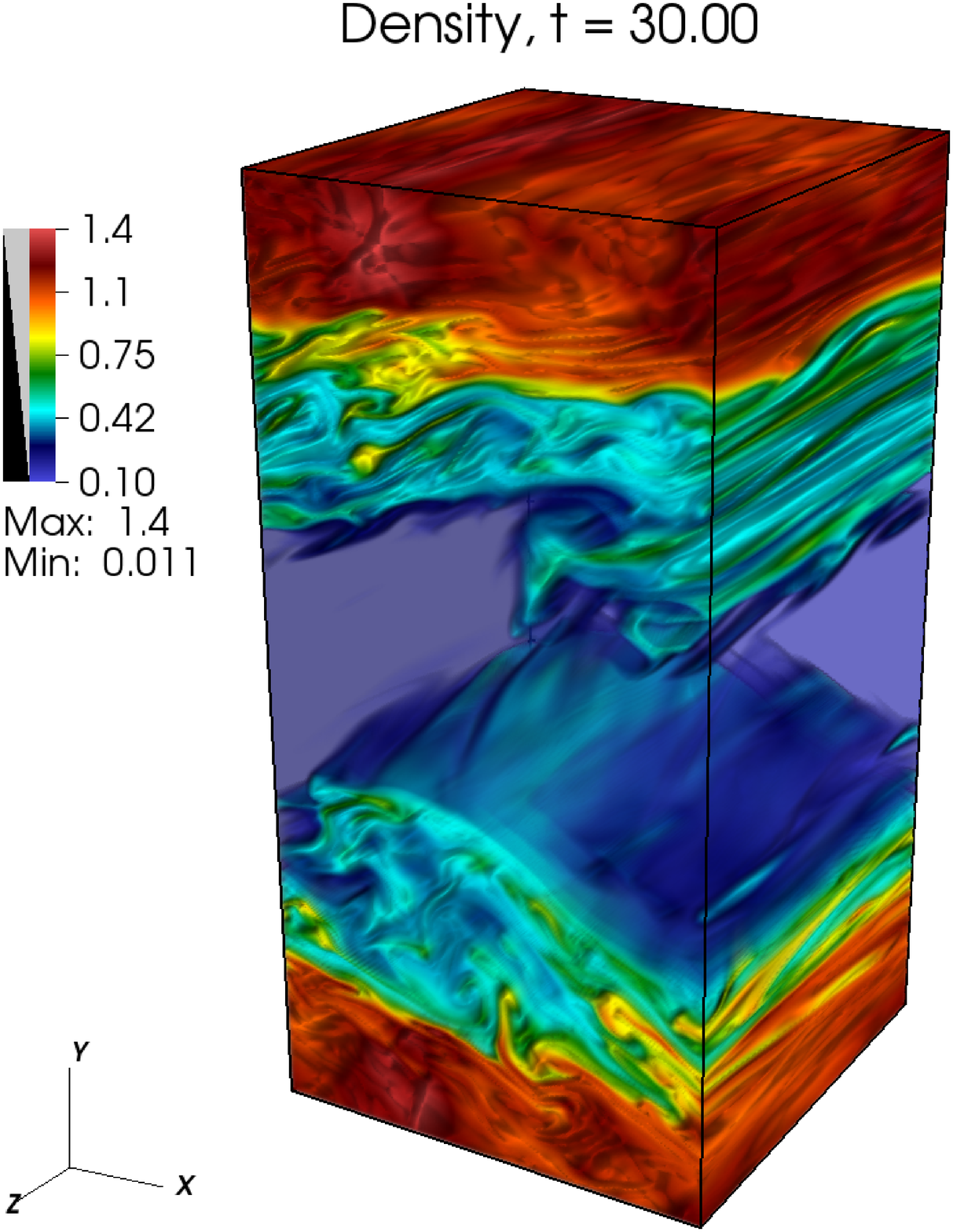}  \\ \vspace{20pt}
  \includegraphics[width=0.3\textwidth]{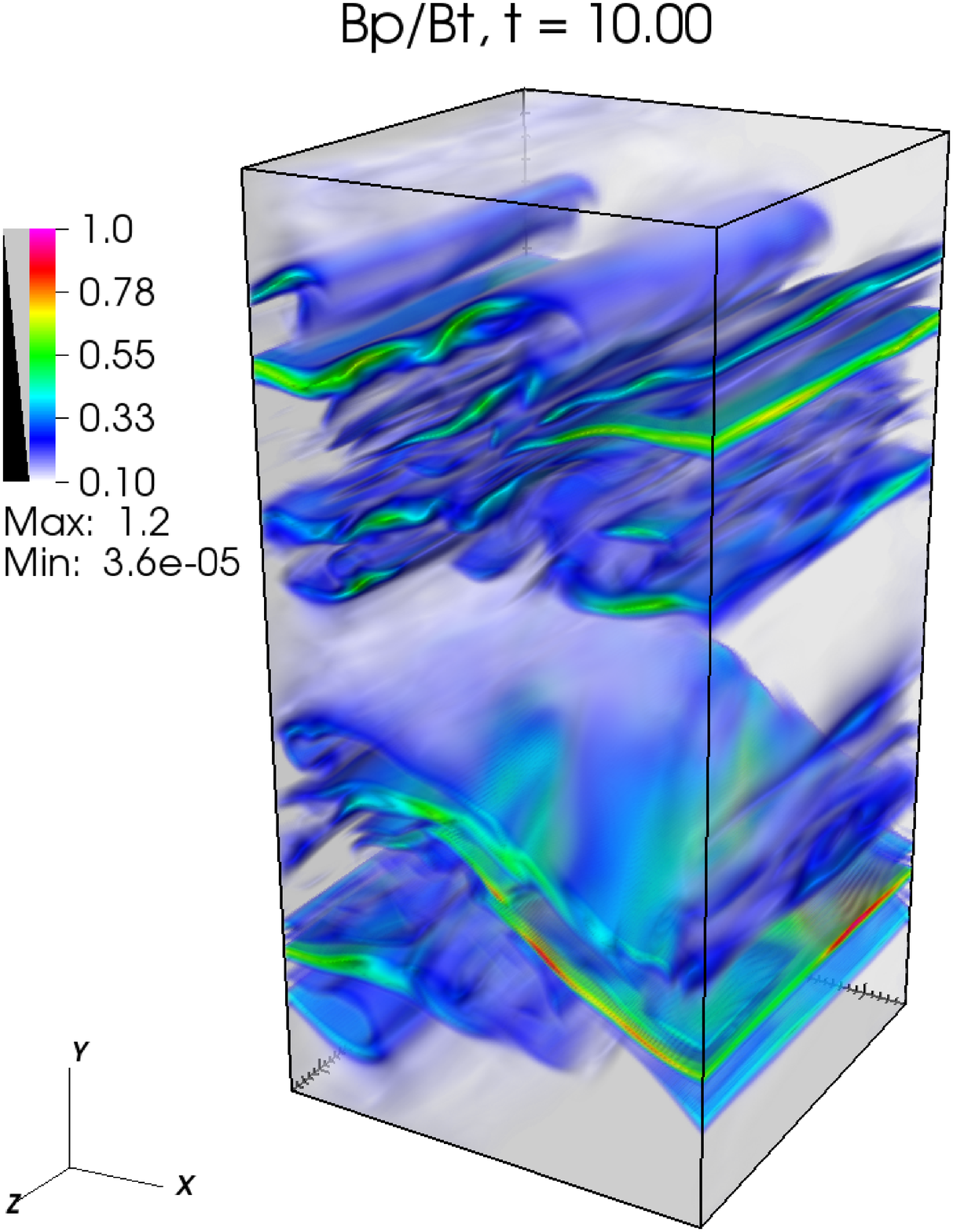} \hspace{10pt}
  \includegraphics[width=0.3\textwidth]{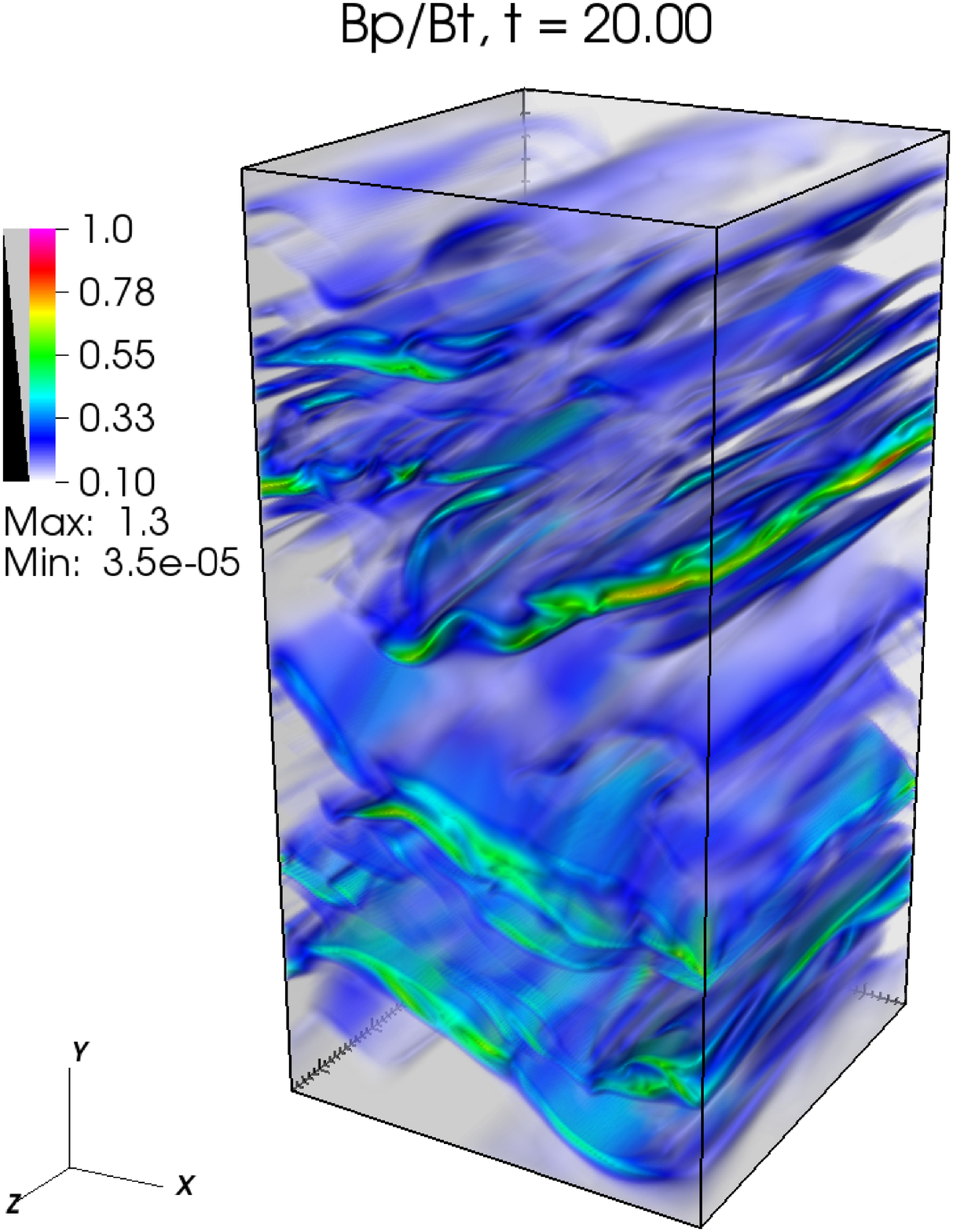} \hspace{10pt}
  \includegraphics[width=0.3\textwidth]{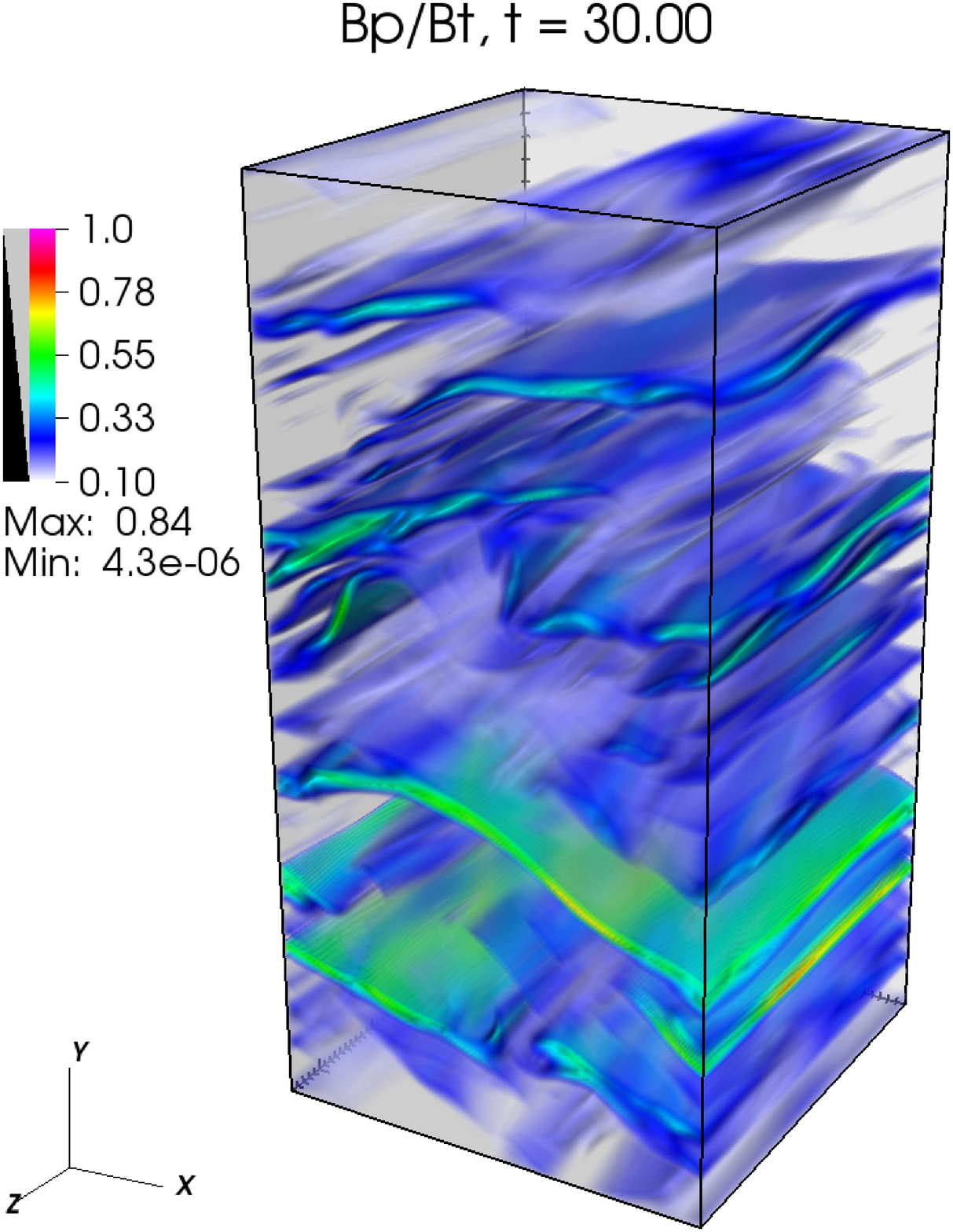}
  \caption{Volume rendering of the density (top) and poloidal to toroidal
           magnetic field (bottom) for the three-dimensional Kelvin-Helmholtz
           instability at $t=10,20$ and $t=30$.
           The conductivity is $\sigma=10^5$ and the MHLLC solver has been employed.}
  \label{fig:kh3D_maps}
\end{figure*}

We extend the previous configuration to three-dimensions by choosing \citep[as in][]{BeckStone2011} the computational domain $x,z\in[-1/2,1/2]$ and $y\in[-1,1]$, using a uniform grid of $256\times512\times256$ zones.
We employ the same initial condition with the exception of the $z$-component of velocity which is now prescribed as in Eq. (\ref{eq:kh:vy}) with the $\sin()$ function replaced by a random number distribution in the range $[-1,1]$.
In order to assess the robustness of our algorithm we employ the most stringent value of $\sigma = 10^5$ using the CT scheme with the MHLLC and the LF solvers.
The system is evolved until $t=30$.

The growth rates, computed again as $\Delta v_y = (\max(v_y)-\min(v_y)/2$, are plotted in the top panel of Fig. \ref{fig:kh3D_growth} for the two Riemann solvers.
In analogy to the 2D case, we observe a faster growth and an earlier transition to the nonlinear regime when the MHLLC solver is employed.

The nonlinear evolution is illustrated in Fig. \ref{fig:kh3D_maps} through a series of volume rendering of density (top) and poloidal to toroidal magnetic field (bottom) at $t=10,20$ and $t=30$.
The large-scale motion remains confined along the initial shear direction and sheet-like thin structures, where most of the magnetic field energy is trapped, characterize the turbulent state.  
These rounded slabs remain roughly parallel to the $z$-direction although significant medium-scale structures develop in this direction for $t\gtrsim20$ around the shear layer.

In order to quantify the numerical diffusion of the two different solvers (which affects the amount of turbulent structure), we evaluate the spectral energy density as
\begin{equation}\label{eq:FFT_power}
  P(k_x) = \int |\mathscr{F}_x(k_x,y,z)|^2 \, dy\,dz \,,\quad
\end{equation}
where $\mathscr{F}_x(k_x, y, z)$ is the one-dimensional fast Fourier transform of the density taken across the $x$ dimension.
Likewise, we construct $P(k_y)$ and $P(k_z)$ by index permutation of (\ref{eq:FFT_power}).
The spectral densities across the three directions are plotted in the bottom panel of Fig. \ref{fig:kh3D_growth} for the MHLLC (blue lines) and the LF (red lines) solvers using solid, dashed and dashed-dotted lines corresponding, respectively, to $P(k_x)$, $P(k_y)$ and $P(k_z)$.
Overall we see that power spectra in the MHLLC case are systematically larger by $\sim 2$ orders of magnitudes (at large wavenumbers) in the $x$- and $y$-directions when compared to the LF case.
While most small-scale power resides in these two directions, the same trend is also observed in turbulent structures across the $z$-direction which make a  significant contribution at large-moderate wavelengths.
Our results favourably compare to those of \cite{BeckStone2011} where the HLLD and HLL solvers have been compared on the same test problem.

While simulations with the GLM scheme could not be completed for such a small value of the resistivity (if not by considerably lowering the Courant number), the results presented in this section demonstrates the our CT scheme is remarkably more stable and robust also in the three-dimensional case.
The computational performance of our CT method yielded a $\sim 13\%$ additional overhead when compared to the same calculation with GLM method (with lower value of $\sigma$).
A comparable performance was obtained in the 3D Blast wave problem, \S\ref{sec:Blast3D}.

\section{Summary}
\label{sec:summary}
%
%
%

In this work we have presented a new second-order Godunov-type constrained transport method for the solution of the resistive relativistic MHD equations in the context of IMplicit-EXplicit (IMEX) Runge-Kutta methods.
Our method follows the work of \cite{Balsara_etal2016} and sets the primary components of electric and magnetic fields at zone interfaces.
A discrete version of Stoke's theorem is employed for the evolution of the electromagnetic fields while hydrodynamic variables are instead located at the zone-center and are treated in the usual finite-volume sense.

During the explicit stages of the IMEX time-stepping, numerical fluxes needed for the update of electromagnetic fields are obtained by applying a two-dimensional Maxwell solver at zone edges  while a standard one-dimensional Riemann solver is used at zone interfaces to advance zone-centered hydrodynamic variables.
This introduces proper upwinding and it ensures that Faraday's law for the the magnetic field is advanced in a divergence-free fashion while the ensuing discretization for the electric field conserves charge to machine precision.
When dealing with the stiff source term, a solution approach for the implicit update of staggered electric field at zone faces that retains the point-local character has been developed.
This has been shown to be formally equivalent to a conservative scheme in which charge is upwinded using a local Lax-Friedrichs flux.
The proposed method of discretization is consistent with Ampere's law from which charge conservation directly follows at the continuous level.

In addition, we have also introduced a new Riemann solver based on the frozen condition of the underlying hyperbolic system of conservation laws with stiff relaxation source terms.
Owing to the weak coupling between Maxwell's and hydrodynamics equations inherent in this limit, the solution to the Riemann problem can be approached by the combination of an outer solver for the EM waves and an inner solver for resolving hydrodynamic waves.
Our composite MHLLC Riemann solver (where \quotes{M} denotes the outer Maxwell while HLLC is the Harten-Lax-van Leer of \cite{MB2005} applied to the hydrodynamic equations) has reduced numerical diffusion when compared to the traditional Lax-Friedrichs (or HLL) solver and it shares a solution procedure analogous to the one outlined in the appendix of \cite{Miranda_etal2018}.

An extensive suite of two- and three-dimensional numerical benchmarks with some new analytical solutions has been used to assess the performance of the newly proposed CT method.
A direct comparison with the widespread GLM scheme of \cite{Palenzuela_etal2009}, that employs a conservative cell-centered discretization, reveals that our CT scheme gives comparable (albeit less diffusive) results for moderate or large resistivities although its benefits are more evident in the ideal limit - small value of the resistivity - in problems where a net charge is produced.
In this regime, we have found that our CT scheme is markedly more robust than the GLM method which instead fails when the time step becomes smaller than the resistivity owing to large spurious oscillations in the charge.
We argue that this may result from the (unstable) explicit discretization of the charge equation used in the standard GLM formalism, where the divergence of the current introduces stiffness.
In support of this, we point out that variants of the GLM scheme in which the charge is not an evolutionary equation and it is computed directly from the divergence of the electric fields \cite[e.g.][]{Dionys_etal2013} do not seem to suffer from this behavior although more investigation is certainly needed.
Similar conclusions can be drawn for the CT scheme of \cite{BdZ2013}, in which the electric field retain a zone-centered representation.

Finally, we point out that spurious (local) charge production may occur at the numerical level in both schemes, owing to discretization errors introduced when taking the divergence of the current.
Most likely, these issues can be ameliorated by introducing schemes with spatial order of accuracy greater than two.
Higher-order reconstruction (such as WENO or PPM) can be easily accommodated for in our formulation although we postpone genuinely third (or higher) finite volume schemes to forthcoming works.

\section*{Acknowledgements}
The authors acknowledge support from the PRIN-MIUR project \emph{Multi-scale
Simulations of High-Energy Astrophysical Plasmas} (Prot.~2015L5EE2Y).
We would also like to thank the referee for his/her constructive
 comments which helped to improve the quality of this manuscript.
\bibliographystyle{mnras}
\bibliography{paper}

\appendix
\section{On the Rest Frame Charge Density}
\label{app:q0}
%
%

We illustrate here the relationship between the vorticity of the flow and the charge density, using the covariant formalism. The \emph{kinematic vorticity} four-vector is defined as \citep{RZ_Book2013}
\begin{equation}
\omega^\lambda = \epsilon^{\mu\nu\lambda\kappa} \nabla_\mu u_\nu \, u_\kappa =
 \epsilon^{\mu\nu\lambda\kappa} \partial_\mu u_\nu \, u_\kappa ,
\end{equation}
where clearly $\omega^\mu u_\mu = 0$.
Thanks to this definition, the covariant derivative of the fluid velocity can be split as
\begin{equation}
  \nabla_\mu  u_\nu = - u_\mu a_\nu + \tfrac{1}{2} \epsilon_{\mu\nu\lambda\kappa}
  \omega^\lambda u^\kappa ,
\end{equation}
where $a^\mu = (u^\nu\nabla_\nu) u^\mu$ is the acceleration, normal to $u^\mu$ too.
Using the definitions of the comoving electromagnetic fields in Eq. (\ref{eq:ebdef}) we derive the following relation
\begin{equation}
  F^{\mu\nu} \nabla_\mu  u_\nu =  e^\mu a_\mu + b^\mu \omega_\mu  .
\end{equation}
Let us now take the divergence of the comoving electric field. Using Maxwell's equations we write
\begin{equation}
  \nabla_\mu e^\mu = \nabla_\mu (F^{\mu\nu}  u_\nu) =  - J^\mu u_\mu + F^{\mu\nu}
  \nabla_\mu  u_\nu ,
\end{equation}
hence, recalling the definition of the comoving charge density $q_0$, we find
\begin{equation}\label{eq:q0}
  q_0  =  \nabla_\mu e^\mu - e^\mu a_\mu  - b^\mu \omega_\mu,
\end{equation}
which differs from the usual Gauss' law.

The above equation provides a link between the evolution of the comoving charge and the electromagnetic fields.
Notice that in the ideal limit (or for small values of the resistivity) the terms with the (comoving) electric field $e^\mu=\eta j^\mu$ are negligible compared to the one with the magnetic field.
Hence in this case we find the simple relation
\begin{equation}
q_0  =  - b^\mu \omega_\mu  ,
\end{equation}
providing $q_0$ directly and involving the kinematic vorticity and the magnetic field alone.
The above scalar product is a relativistic invariant, therefore it is convenient to calculate it in the comoving frame of the fluid, that is, for a flat spacetime metric
\begin{equation}
  u^\mu = (1,\vec{0}), \quad \omega^\mu = ( 0 , \boldsymbol\omega), \quad b^\mu = (0,
\vec{B}),
\end{equation}
%
%
where $\boldsymbol\omega = \vec{\nabla}\times \vec{u}$ is the vorticity three-vector
(recall that the velocity vanishes but not its spatial derivatives). The comoving
charge density then becomes, in this case
\begin{equation}\label{eq:omegaB}
  q_0 = - \vec{B}\cdot\boldsymbol\omega .
\end{equation}

\section{Jacobian of the IMEX-Newton Method}
\label{app:NB_Jacobian}
%
%
%

In order to apply the Newton-Broyden scheme during the implicit stage of our SSP-IMEX scheme, the Jacobian (\ref{eq:f_NB}) must be computed:
\begin{equation}\label{eq:Jacobian_app}
 \tens{J}_{ij} = -Du_i\pd{h(\vec{u})}{u_j}
                 - \varepsilon_{ikl}\pd{E_k(\vec{u})}{u_j}B_{l}
                 - Dh(\vec{u})\delta_{ij} \,.
\end{equation}
where $\varepsilon_{ijk}$ is the Levi-Civita symbol.

For the ideal equation of state (\ref{eq:eos}) the gradient of the specific enthalpy $h=w/\rho$ is obtained as:
\begin{equation}
 \pd{h}{u_i} = \frac{\Gamma_1}{D}\left(\pd{p}{u_i} + \frac{p}{\gamma}u_i\right) \,,
\end{equation}
while the gradient of the pressure is obtained by differentiating the second in (\ref{eq:recprimitive}) while keeping ${\cal E}$ and $\vec{B}$ constant:
\begin{equation}
 \pd{p}{u_i} = -\frac{1}{\Gamma_1\gamma^2-1} \left(
                 2u_i\Gamma_1 p +
                \frac{Du_i}{\gamma} + \pd{E^2}{u_i}\right) \,.
\end{equation}

The second term on the right hand side of Eq. (\ref{eq:Jacobian_app}) involves derivatives of the electric field.
Using Eqns. (\ref{eq:implicitE_solve}) and (\ref{eq:Eu}), we split them as the sum of three terms:
\begin{equation}
 \vec{E} = \eta{\cal A}\vec{R}^{(s-1)}_e + \vec{H} + \vec{K} \,,
\end{equation}
where
\begin{equation}\label{eq:AHK}
 \left\{
 \begin{array}{lcl}
  {\cal A} = \DS\frac{1}{\eta + \delta t\gamma} \\ \noalign{\medskip}
  \vec{H} = -{\cal A}\delta t(\vec{u}\times\vec{B}) \\ \noalign{\medskip}
  \vec{K} = {\cal A}\left(\DS\frac{\delta t\eta\vec{R}_e^{(s-1)}\cdot\vec{u}}{\eta\gamma + \delta t}\right)\vec{u} \,.
 \end{array}\right.
\end{equation}
We proceed by first computing the derivatives of ${\cal A}$ which is a scalar function:
\begin{equation}
 \pd{\cal A}{u_i} = -{\cal A}^2\delta t\frac{u_i}{\gamma} \,.
\end{equation}
Next we differentiate $\vec{H}$:
\begin{equation}
 \pd{H_i}{u_j} = -\delta t(\vec{u}\times\vec{B})_i\pd{\cal A}{u_j} - \delta t{\cal A}\varepsilon_{ijk}B_k,
\end{equation}

To obtain the derivatives of the last term ($\vec{K}$) in Eq. (\ref{eq:AHK}) we first define the quantity 
\begin{equation}
 {\cal D} = \frac{\vec{R}_e^{(s-1)}\cdot\vec{u}}{\gamma\eta + \delta t} \,,
\end{equation}
together with its derivatives,
\begin{equation}
 \pd{\cal D}{u_i} =   \frac{R^{(s-1)}_{e,i}}{\gamma\eta + \delta t}
                    - \frac{\eta u_i}{\gamma}
                      \frac{\vec{R}^{(s-1)}_e\cdot\vec{u}}{\gamma\eta + \delta t}\,,
\end{equation}
so that the gradient of $\vec{K}$ with respect to velocity is calculated as
\begin{equation}
 \pd{K_i}{u_j} =   u_i\eta\delta t{\cal D}\pd{\cal A}{u_j}
                 + u_i\eta\delta t{\cal A}\pd{\cal D}{u_i}
                 + \delta_{ij}\eta\delta t{\cal A}{\cal D} \, .
\end{equation}
Putting all toegther, the derivatives of the electric field are finally given by 
\begin{equation}
 \pd{E_i}{u_j} =   \eta R^{(s-1)}_{e,i}\pd{\cal A}{u_j}
                 + \pd{H_i}{u_j}
                 + \pd{K_i}{u_j}\,.
\end{equation}

\bsp	
\label{lastpage}
\end{document}